\documentclass[ amsmath,amssymb,reprint,onecolumn,longbibliography,
 aps]{revtex4-1}

\usepackage{graphicx}
\usepackage{dcolumn}
\usepackage{bm}
\usepackage{amsmath}
\usepackage{amsfonts}
\usepackage[colorlinks=true, citecolor=red, linkcolor=blue,urlcolor=blue ]{hyperref}

\hypersetup
{
	pdfauthor={Weida Liao and Eric Lauga},
	pdftitle={Energetics of synchronisation for model flagella and cilia},
}

\begin{document}

\title{Energetics of synchronisation for model flagella and cilia}

\date{\today}
\author{Weida Liao}
\author{Eric Lauga}
\email{e.lauga@damtp.cam.ac.uk}
\affiliation{Department of Applied Mathematics and Theoretical Physics, University of Cambridge, Wilberforce Road, CB3 0WA, Cambridge, UK}

\begin{abstract}

Synchronisation is often observed in the swimming of flagellated  cells, either for multiple appendages on the same organism or between the flagella of nearby cells. Beating cilia are also seen to synchronise their dynamics. 
In 1951,  Taylor  showed that the observed in-phase beating of the flagella of co-swimming spermatozoa was consistent with minimisation of the  energy dissipated in the surrounding fluid.  Here we revisit Taylor's hypothesis for three   models of flagella and cilia: 
(1) Taylor's waving sheets with both longitudinal and transverse  modes, as relevant for flexible flagella; 
(2) spheres orbiting above a no-slip surface to model  interacting flexible cilia;  and 
(3) whirling rods above a no-slip surface to address the interaction of nodal cilia. 
By calculating the flow fields explicitly, we show that the rate of working of the model flagella or cilia is minimised in our three models for (1) a phase difference depending on the separation of the sheets and precise waving kinematics; (2) in-phase or opposite-phase motion depending on the relative position and orientation of the spheres; and (3) in-phase whirling of the rods. These results will be useful in future  models probing the dynamics of synchronisation in these setups.

\end{abstract}
	\maketitle

\section{Introduction}

The  majority of  cells able to move in fluids do so with the aid of slender cellular appendages called flagella or cilia~\cite{braybook}, whose time-varying motion generates  hydrodynamic stresses that propel the cells forward~\cite{Lighthill76,brennen77,childress81,Lauga09,laugabook}. 
Although different organisms employ different methods, the same general physical principles of swimming at low Reynolds number apply to the locomotion of all cells, from simple bacteria~\cite{lauga16} to spermatozoa~\cite{fauci06,gaffney11} and higher aquatic microorganisms~\cite{pedley92,stocker}.

One of the peculiar features of swimming cells is their ability to synchronise the motion of their appendages. 
Indeed, synchronisation is observed to take place either for  organisms equipped with multiple flagella or cilia, or between the flagella of nearby cells~\cite{lg12}. 
Such synchronisation occurs, for example, for  bacteria equipped with multiple helical flagella~\cite{Kim03,Kim2004,Reichert05,Janssen11,Reigh12,Reigh13}. 
Spermatozoa  swimming close to one another are also known to cooperate~\cite{hayashi98} and  synchronise their waving motion~\cite{yang08,woolley09}, even though each cell is   independently actuated. 
A unicellular alga can synchronise its two flagella~\cite{goldstein2009noise}, while hydrodynamic interactions also lead to synchronised dynamics for appendages  on different cells~\cite{brumley2014flagellar}. 
For larger ciliated cells~\cite{blake74}, or for
multicellular aquatic organisms equipped with many short flagella~\cite{brumley2012}, the synchronisation dynamics takes the form of 
metachronal waves, akin to spectator waves in sport  stadiums. Synchronisation also extends beyond swimming to the dynamics of cilia on biological tissues. Nodal cilia, which rotate rigidly around tilted conical orbits and produce left--right asymmetry in embryos, are a famous example of this~\cite{okada2005mechanism,cartwright2004fluid}.

While the synchronised dynamics can often  be described mathematically within the framework of  coupled oscillators~\cite{strogatz00}, a  key question lies in identifying the physical ingredients leading to synchronisation. 
In order for two oscillators to synchronise, it is intuitive that two features are required. 
First, there needs to be a physical means for the oscillators to communicate. 
It has long been suspected that this is generically achieved for swimming cells through hydrodynamic interactions~\cite{goldstein16}, though ``dry" synchronisation  is also possible via coupling to the cell body~\cite{bennett2013,geyer2013cell} or  	intracellular coupling~\cite{quaranta2015,wan2016coordinated}.

The second required ingredient is a physical mechanism that allows the phase of the oscillators to evolve. 
In the context of cellular synchronisation via hydrodynamic interactions, two  such mechanisms have been identified theoretically. 
In the first one, it is the    elastic compliance of the orbits that allows oscillators to speed up or slow down in response to hydrodynamic flows~\cite{niedermayer08}. 
The second generic mechanism is phase-dependent cellular forcing, so that flagella or cilia are able to change their phase dynamics in response to external hydrodynamic stresses~\cite{uchida2011}.

One interesting aspect of cellular synchronisation concerns its energetic consequences. 
The degrees of freedom available to interacting flagella and cilia could be used  to minimise energetic costs, for example, the dissipation of mechanical  energy in the surrounding viscous fluid. 
This was the argument originally examined by Taylor to explain the  in-phase synchronisation of nearby spermatozoa~\cite{taylor51}. While this energetic view has proven to be popular in addressing the collective dynamics of cilia~\cite{gueron99} and metachronal waves~\cite{michelin10}, theoretical calculations have shown that the dynamics of swimmers can   cause cells to synchronise into a state where the energy dissipation is maximum~\cite{elfring09}.  
The states of minimum energy also depend critically on the relative position and orientation of the flagella~\cite{mettot2011energetics}.

In this paper, we  study the relationship between synchronised dynamics and the  dissipation in the surrounding fluid using three  classical    models  for  flagella and cilia. 
We start in Sec.~\ref{sec:sheets}  by revisiting Taylor's two-dimensional swimming sheet model, relevant to the synchronisation of spermatozoa~\cite{taylor51,elfring09,elfring2011passive,elfring2010two,elfring11}, and extending it to the general case of flexible swimmers undergoing both longitudinal and transverse waving.
In this case, we show that the minimum of energy dissipation does not necessarily occur at the in-phase configuration but  at an optimal phase difference between the two sheets that depends on the detailed kinematics of the sheets. We further use a long-wavelength  calculation to provide  physical intuition  in the case of swimmers with purely longitudinal waves.  
In Sec.~\ref{sec:spheres}, we then consider orbiting spheres, a  minimal  model   used to address the dynamics of interacting cilia in past theoretical~\cite{lagomarsino02,lagomarsino03,lenz06,vilfan06,niedermayer08} and experimental work~\cite{kotar2010hydrodynamic,kotar2013optimal}. 
We find that depending on the orientations and relative positions of the orbits, either in-phase or opposite-phase motion of the spheres minimises energy dissipated. 
In Sec.~\ref{sec:rods}, we finally consider interacting nodal cilia~\cite{okada2005mechanism} modelled by  rods whirling above a no-slip surface~\cite{smith2011mathematical,cartwright2004fluid}. 
In this case, the energy is seen to be always minimised at the in-phase configuration. 
We finish  in Sec.~\ref{sec:discussion} with a discussion of our results in the context of the dynamics of biological synchronisation.
Our work, able to characterise analytically the states of minimum dissipated energy in these  models, will be useful for future work investigating the complex synchronised dynamics of realistic flagella and cilia.

\section{Model for interacting flagella: Two waving sheets}\label{sec:sheets}

\subsection{Setup}\label{sec:smallamplsetup}
\begin{figure}
	\includegraphics[width=0.7\textwidth]{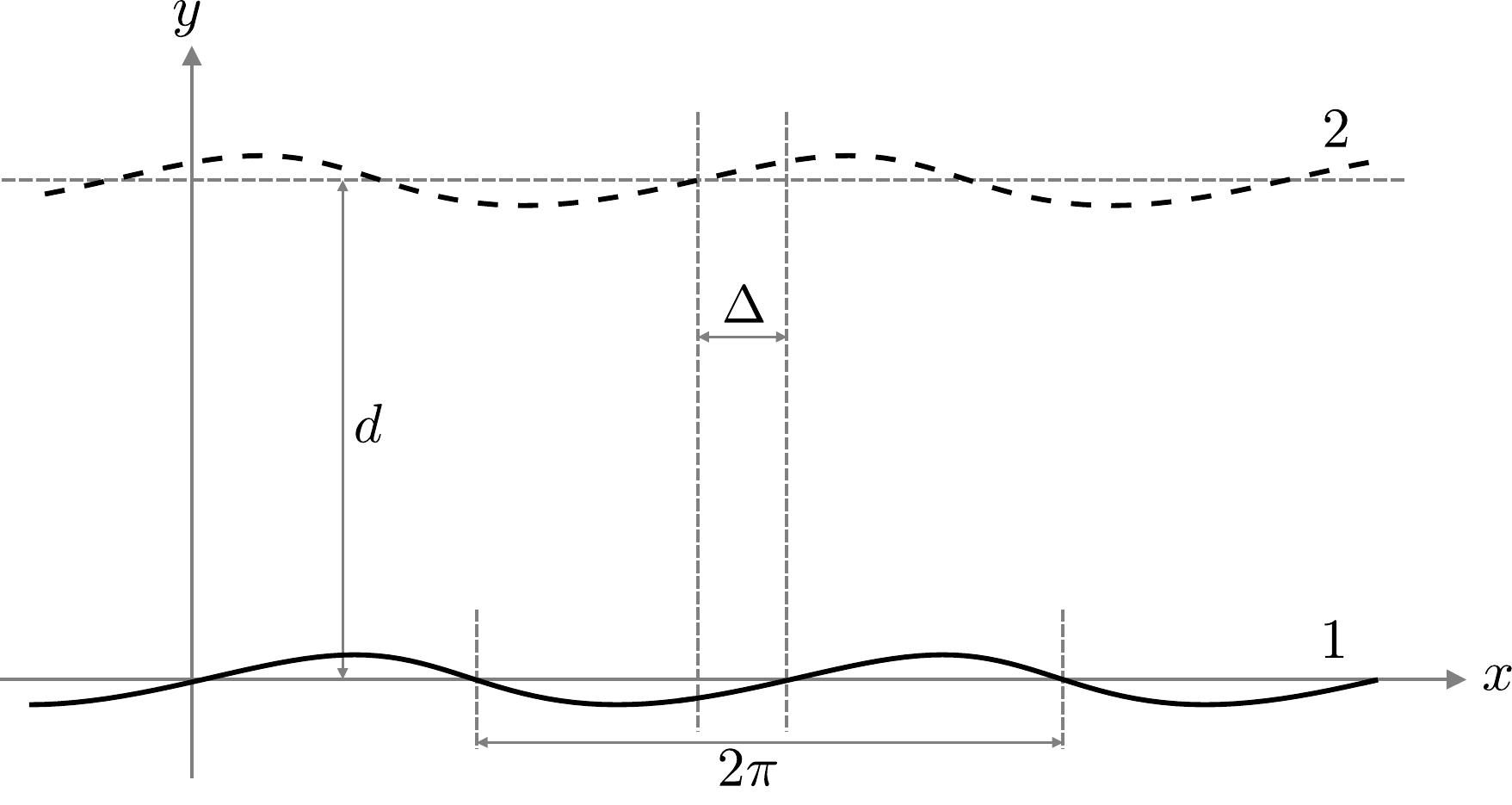}
	\caption{Two identical waving sheets as models for two flagellated spermatozoa, which interact hydrodynamically. The sheets, of  dimensionless period $2\pi$,  are separated by a mean dimensionless distance $d$. The phase difference between the sheets is denoted by $\Delta$. The waving motion includes both  longitudinal and transverse modes, so that a material point traces an ellipse.}
	\label{fig:diagram_sheets}
\end{figure}

In this first section, we consider the hydrodynamics of two waving sheets as models for two identical flagellated spermatozoa, which interact through a viscous fluid. 
The setup is illustrated in Fig.~\ref{fig:diagram_sheets} in dimensionless form.
The sheets are infinite and two-dimensional, as in Taylor's original article~\cite{taylor51}.
The two sheets are separated by a mean distance $h$, the first at the unperturbed position $y=0$ and the second at unperturbed position $y=h$.
The two sheets undergo identical prescribed waving motion but with an imposed constant phase difference $\Delta$.  
The waving motion includes both longitudinal and transverse deformation modes with   amplitudes $A$ and $B$ respectively.
For each swimmer, we denote by $\phi$ a phase difference between these two modes, where $\phi=0$ means that the material points move in ellipses with semi-axes parallel to the $x$ and $y$ axes.
The shape of each sheet is periodic in space with wavenumber $k$ and in time with angular frequency $\omega$.  
The material points on the first sheet are given by $\left (x_\text{s}^{(1)},y_\text{s}^{(1)}\right )$, with similar notation for the second sheet.

To proceed with the calculation, we will consider the limit of small-amplitude waving and solve for the leading-order rate at which work is done by the sheets on the fluid between them, as a function of the phase difference $\Delta$. (The work done below sheet 1 and above sheet 2 is independent of the phase difference and therefore we do not need to compute it.) We will compute the optimal phase difference $\Delta^*$, which minimises this rate of working with respect to $\Delta$.  
Taylor~\cite{taylor51} previously found that if the sheets undergo only transverse deformation ($A=0$), then the rate of working at leading order is always minimised by in-phase waving of the sheets ($\Delta^*=0$).
In our paper, we allow the sheets to be flexible and thus undergo both longitudinal and transverse deformation; that is, both $A$ and $B$ may be nonzero. For this setup, we then find $\Delta^*$ as a function of the amplitudes, mean sheet separation and phase difference between the modes.

We impose the kinematics of material points with unperturbed position $x$ on each sheet at time $t$ as
\begin{align}
    x_\text{s}^{(1)} &= x + A \cos(kx-\omega t-\phi), \\
    y_\text{s}^{(1)} &= B \sin(kx- \omega t), \\
    x_\text{s}^{(2)} &= x + A \cos(kx-\omega t+\Delta-\phi), \\
    y_\text{s}^{(2)} &= h + B \sin(kx-\omega t+\Delta).
\end{align}
Thus, a material point traces out an ellipse in time. In the case of longitudinal oscillation only or transverse oscillation only of each material point, this ellipse becomes a line segment. If there is no longitudinal oscillation, then the material point has position $x_\text{s}=x$.
 
We note that the phase difference $\Delta$ is defined only up to a multiple of $2\pi$. 
In our later discussion, we will refer to values of $\Delta$ that satisfy $-\pi\leq\Delta<\pi$. 
Thus, if $\Delta$ is small and positive, then the phase of the second sheet is slightly ahead of that of the first sheet; that is, the peaks of the second sheet  are slightly to the left of those of the first sheet. 
Conversely, if $\Delta$ is small and negative, then the phase of the second sheet is slightly behind that of the first sheet, and the peaks of the second sheet are slightly to the right of those of the first sheet.

The   fluid between the sheets is assumed to be Newtonian and the Reynolds number sufficiently small that the  governing equations  are the incompressible Stokes equations 
\begin{align}
\nabla\cdot\boldsymbol{\sigma}= \mu\nabla^{2}\mathbf{u} -\nabla p&=\mathbf{0},\label{eq:Stokes}\\
\nabla\cdot\mathbf{u}&=0\label{eq:incompressible},
\end{align}
where $\boldsymbol{\sigma}$ is the stress tensor, $\mathbf{u}$ is the fluid velocity field, $p$ is the dynamic pressure and $\mu$ is the dynamic viscosity~\cite{kimbook}. The stress tensor $\boldsymbol{\sigma}$ is given by $\boldsymbol{\sigma}=-p\mathbf{1}+\mu\left [\nabla\mathbf{u}+(\nabla\mathbf{u})^T\right ]$, where $\mathbf{1}$ is the identity tensor.

We nondimensionalise the problem using $k^{-1}$ and $\omega^{-1}$ as characteristic length and time scales, so that the pressure scale is $\mu \omega$. 
The sheets are then separated by a mean dimensionless distance $d\equiv kh$ (see Fig.~\ref{fig:diagram_sheets}).
In preparation for calculation of the flow in the small-amplitude limit, we define the small parameter $\epsilon \equiv \left[(Ak)^2+(Bk)^2\right]^{\frac{1}{2}}$, so that we have dimensionless longitudinal and transverse amplitudes $a\equiv{Ak}/{\epsilon}$ and $b\equiv{Bk}/{\epsilon}$ respectively. 
We now view $a$ and $b$ as independent of $\epsilon$ and let $x, y, t, \mathbf{u}, p$ and $\boldsymbol{\sigma}$ denote their corresponding dimensionless quantities. 
Then the dimensionless kinematics for the sheets become
\begin{align}
    x_\text{s}^{(1)} &= x + \epsilon a \cos(x-t-\phi), \label{eq:sheetsdimlesskinx1}\\
    y_\text{s}^{(1)} &= \epsilon b \sin(x-t), \label{eq:sheetsdimlesskiny1}\\
    x_\text{s}^{(2)} &= x + \epsilon a \cos(x-t+\Delta-\phi),\label{eq:sheetsdimlesskinx2} \\
    y_\text{s}^{(2)} &= d + \epsilon b \sin(x-t+\Delta).\label{eq:sheetsdimlesskiny2}
\end{align}
Note that we use the shared swimming frame of the two sheets, which has coordinates $(x,y)$. 
This is possible at order $\epsilon$ by the following symmetry argument. 
A replacement $\epsilon\mapsto -\epsilon$ is equivalent to a translation of the setup by half a wavelength in the $x$ direction. 
Such a translation does not affect the relative translational velocity of the sheets, because the sheets are infinite and periodic in space. 
Hence, the relative translational velocity is even in $\epsilon$.
Therefore, the order $\epsilon$ relative translational velocity of the sheets is zero.

The no-slip velocity boundary conditions for flow on the sheets are given by
\begin{align}
\left .u_x\right \vert_{\left(x_\text{s}^{(1)}, y_\text{s}^{(1)}\right)} &= \epsilon a \sin(x-t-\phi),\\
\left .u_y\right \vert_{\left(x_\text{s}^{(1)}, y_\text{s}^{(1)}\right)} &= -\epsilon b \cos(x-t),\\
\left .u_x\right \vert_{\left(x_\text{s}^{(2)}, y_\text{s}^{(2)}\right)} &= \epsilon a \sin(x-t+\Delta-\phi),\\
\left. u_y\right \vert_{\left(x_\text{s}^{(2)}, y_\text{s}^{(2)}\right)} &= -\epsilon b \cos(x-t+\Delta).
\end{align}

Since the problem is two-dimensional, we may introduce a stream function $\psi$, which enforces incompressibility of the flow, with
\begin{align}
    u_x&= \frac{\partial \psi}{\partial y},\\ u_y&=-\frac{\partial \psi}{\partial x}.
\end{align}
The Stokes equations imply classically that the stream function $\psi$ is a solution to the biharmonic equation~\cite{happel}
\begin{align}
\nabla^4\psi=0.
\end{align}
Written using this stream function, the velocity boundary conditions for flow on the sheets then become
\begin{align}
    \left .\frac{\partial \psi}{\partial x}\right \vert_{\left(x_\text{s}^{(1)}, y_\text{s}^{(1)}\right)} &= \epsilon b \cos(x-t),\\
    \left .\frac{\partial \psi}{\partial y}\right \vert_{\left(x_\text{s}^{(1)}, y_\text{s}^{(1)}\right)} &= \epsilon a \sin(x-t-\phi),\\
    \left. \frac{\partial \psi}{\partial x}\right \vert_{\left(x_\text{s}^{(2)}, y_\text{s}^{(2)}\right)} &= \epsilon b \cos(x-t+\Delta),\\
    \left .\frac{\partial \psi}{\partial y}\right \vert_{\left(x_\text{s}^{(2)}, y_\text{s}^{(2)}\right)} &= \epsilon a \sin(x-t+\Delta-\phi).
\end{align}
These boundary conditions are nonlinear in $\epsilon$. We therefore use an asymptotic approach and consider in what follows the limit of small amplitude. 
Since there is no flow if $\epsilon=0$, we  may use the following regular perturbation expansions in $\epsilon$ of the stream function $\psi$, fluid velocity field $\mathbf{u}$, pressure $p$ and stress tensor $\boldsymbol{\sigma}$
\begin{align}
    \psi &= \epsilon \psi_1 + O \left (\epsilon^2\right ),\\
    \mathbf{u}&=\epsilon\mathbf{u}_1+O\left (\epsilon^2\right ),\\
    p&=\epsilon p_1 + O\left (\epsilon^2\right ),\\
    \boldsymbol{\sigma}&=\epsilon\boldsymbol{\sigma}_1+O\left (\epsilon^2\right ).
\end{align}
Using a Taylor expansion of the velocity boundary conditions, we thus obtain the boundary conditions for $\mathbf{u}_1$  on both swimmers as
\begin{align}
\left .u_{1,x}\right \vert_{y=0} &= a \sin(x-t-\phi),\label{eq:sheetsu1x0} \\
\left .u_{1,y}\right \vert_{y=0} &= -b\cos(x-t),\label{eq:sheetsu1y0}\\
\left .u_{1,x}\right \vert_{y=d} &= a \sin(x-t+\Delta-\phi), \label{eq:sheetsu1xd}\\
\left .u_{1,y}\right \vert_{y=d} &= -b\cos(x-t+\Delta).\label{eq:sheetsu1yd}
\end{align}
Note that these order $\epsilon$ boundary conditions are now applied at the unperturbed positions of the two sheets, which  no longer involve the value of $\epsilon$.

\subsection{Flow at order $\epsilon$}

The first-order stream function $\psi_1$ in the fluid between the two sheets satisfies the biharmonic equation
\begin{align}
\nabla^4\psi_1=0.
\end{align}
Since the Stokes equations are linear, we may use complex notation for $\psi_1$ and implicitly take real parts. Then the boundary conditions at order $\epsilon$ in Eqs.~(\ref{eq:sheetsu1x0})--(\ref{eq:sheetsu1yd}) become
\begin{align}
    \left .\frac{\partial \psi_1}{\partial x}\right \vert_{(x, 0)} &= b \exp[i(x-t)],\label{eq:sheetspsi1x0}\\
    \left .\frac{\partial \psi_1}{\partial y}\right \vert_{(x, 0)} &= -ai\exp[i(x-t-\phi)], \label{eq:sheetspsi1y0}\\
    \left .\frac{\partial \psi_1}{\partial x}\right \vert_{(x, d)} &= b \exp[i(x-t+\Delta)],\label{eq:sheetspsi1xd}\\
    \left .\frac{\partial \psi_1}{\partial y}\right \vert_{(x, d)} &= -ai \exp[i(x-t+\Delta-\phi)].\label{eq:sheetspsi1yd}
\end{align}

The general unit-speed, $2\pi$-periodic solution to the biharmonic equation is
\begin{align}
\psi_1 = Ky+Ly^2+My^3+\sum_{n\geq 1}\{(G_n+yH_n)\cosh[n(y-d)] + (I_n+yJ_n)\sinh[n(y-d)]\}\exp[in(x-t)],
\end{align}
where the capital letters denote constants to be determined and we have omitted an arbitrary, physically irrelevant additive constant~\cite{laugabook}.
Since we impose zero mean pressure gradient in the $x$ direction, it is necessary to have  $M=0$. Similarly, asking for  zero mean shear acting on the swimmers means that we also need $L=0$. 
Now, by comparing with the boundary conditions in Eqs.~(\ref{eq:sheetspsi1x0})--(\ref{eq:sheetspsi1yd}), we see that $K=0$ and that only the $n=1$ mode in the infinite sum is nonzero.
Dropping the subscript 1 from the $n=1$ mode coefficients, we write down the relevant solution for $\psi_1$ as
\begin{align}
    \psi_1 = [(G+yH)\cosh(y-d) + (I+yJ)\sinh(y-d)]\exp[i(x-t)].
\end{align}
Substituting this into the boundary conditions, we obtain a system of simultaneous equations
\begin{align}
    i(G \cosh{d} - I \sinh{d}) &= b, \label{eq:sim1}\\
    (H+I) \cosh{d} - (G+J) \sinh{d} &= -ai\exp(-i \phi),\label{eq:sim2} \\
    i(G+dH) &= b \exp(i \Delta),\label{eq:sim3} \\
    H+I+dJ &= -ai\exp[i (\Delta-\phi)].\label{eq:sim4}
\end{align}

\subsection{Leading-order rate of working}

We now consider the energetics of the problem  and calculate the rate at which work is done $\dot{W}$ by the sheets, at leading order in the waving amplitude, on the fluid between them (per wavelength). 
Since the rate of working varies quadratically with the flow velocity, its leading-order value occurs at order $\epsilon^2$ and we may use the regular perturbation expansion 
\begin{align}
	\dot{W}=\epsilon^2\dot{W}_2+O\left (\epsilon^3\right ).
\end{align}
Thus, we only need the stream function to order $\epsilon$. 
Note that work is also done on the fluid \em not \em between the sheets~\cite{childress81,laugabook}. 
However, as noted earlier, this does not depend on the phase difference $\Delta$, so is irrelevant to minimisation with respect to $\Delta$ of the total rate of working. 

The leading-order rate of working $\dot{W}_2$ per wavelength is given by
\begin{align}
    \dot{W}_2 = - \int_{0}^{2\pi} \left .u_{1,x} \sigma_{1, xy} \right \vert_{y=0} \mathrm{\,d} x - \int_{0}^{2\pi} \left .u_{1,y} \sigma_{1, yy} \right \vert_{y=0} \mathrm{\,d} x \nonumber \\
    + \int_{0}^{2\pi} \left .u_{1,x} \sigma_{1, xy} \right \vert_{y=d} \mathrm{\,d} x + \int_{0}^{2\pi} \left .u_{1,y} \sigma_{1, yy} \right \vert_{y=d} \mathrm{\,d} x,
\end{align}
where the relevant components of the stress tensor $\boldsymbol{\sigma}$ are
\begin{align}
\sigma_{1, xy} &= \frac{\partial u_{1,x}}{\partial y} + \frac{\partial u_{1,y}}{\partial x}, \\
\sigma_{1, yy} &= -p_1 + 2\frac{\partial u_{1,y}}{\partial y}.
\end{align}
We find the leading-order pressure by integrating the $x$ component of the Stokes equations at order $\epsilon$
\begin{align}
    \frac{\partial p_1}{\partial x} = \nabla^2 \frac{\partial \psi_1}{\partial y},
\end{align}
which gives
\begin{align}
    p_1 = -2i[H \cosh(y-d) + J \sinh(y-d)] \exp[i(x-t)].
\end{align}
We may then use the  boundary conditions on $\psi_1$ in Eqs.~(\ref{eq:sheetspsi1x0})--(\ref{eq:sheetspsi1yd}) to obtain components of the stress tensor in complex notation 
\begin{align}
    \left .\sigma_{1, xy}\right \vert_{y=0} =& [(G+2J)\cosh{d} - (2H+I)\sinh{d} -ib]\exp[i(x-t)], \\
    \left .\sigma_{1, yy}\right \vert_{y=0} =& \left [2i(H\cosh{d}-J\sinh{d}) - 2a\exp(-i\phi)\right ]\exp[i(x-t)], \\
    \left .\sigma_{1, xy}\right \vert_{y=d} =& \left [G+dH+2J  - ib\exp(i\Delta)\right ]\exp[i(x-t)],\\
    \left .\sigma_{1, yy}\right \vert_{y=d} =& \left \{2iH - 2a\exp[i(\Delta-\phi)]\right \}\exp[i(x-t)].
\end{align}
Using the real and imaginary parts of the simultaneous equations in  Eqs.~(\ref{eq:sim1})--(\ref{eq:sim4}), we may eliminate $G$ and $I$ to obtain in real notation
\begin{align}
    \left .\sigma_{1, xy}\right \vert_{y=0} =& [\operatorname{Re}(G+2J)\cosh{d} -\operatorname{Re}(2H+I)\sinh{d}]\cos(x-t) \nonumber\\   
    &+ [\operatorname{Im}(2H+I)\sinh{d} -\operatorname{Im}(G+2J)\cosh{d} +b]\sin(x-t) \nonumber\\   
    =& 2[\operatorname{Re}(J)\cosh{d} - \operatorname{Re}(H)\sinh{d}]\cos(x-t) \nonumber\\ &+ 2[b-\operatorname{Im}(J)\cosh{d}+\operatorname{Im}(H)\sinh{d}]\sin(x-t), \\
    \left .\sigma_{1, yy}\right \vert_{y=0} =& 2[\operatorname{Re}(J)\sinh{d}-\operatorname{Re}(H)\cosh{d}]\sin(x-t) \nonumber\\
    &+2[\operatorname{Im}(J)\sinh{d}-\operatorname{Im}(H)\cosh{d}]\cos(x-t) -2a\cos(x-t-\phi), \\
    \left .\sigma_{1, xy}\right \vert_{y=d} =& \operatorname{Re}(G+dH+2J)\cos(x-t) - \operatorname{Im}(G+dH+2J)\sin(x-t) + b\sin(x-t+\Delta) \nonumber\\ 
    =& [b\sin{\Delta}+2\operatorname{Re}(J)]\cos(x-t) +[b\cos\Delta -2\operatorname{Im}(J)]\sin(x-t) +b\sin(x-t+\Delta), \\
    \left .\sigma_{1, yy}\right \vert_{y=d} =& -2\operatorname{Re}(H)\sin(x-t) - 2\operatorname{Im}(H)\cos(x-t) - 2a\cos(x-t+\Delta-\phi).
\end{align}
Solving  Eqs.~(\ref{eq:sim1})--(\ref{eq:sim4}) for $H$ and $J$ gives
\begin{align}
H=-\frac{i}{\sinh ^2d-d^2} \big\{&a \sinh ^2d \exp [i (\Delta -\phi )]+a d \sinh d \exp (-i \phi )-b d \exp (i \Delta )\nonumber\\
&-b \sinh d \cosh d \exp (i \Delta )+b \sinh d+b d \cosh d\big\},\\
J=-\frac{i}{\sinh ^2d-d^2} \big\{&-a d \exp [i (\Delta -\phi )]+a \sinh d \cosh d \exp [i (\Delta -\phi )]-a \sinh d \exp (-i \phi )\nonumber\\
&+a d \cosh d \exp (-i \phi )-b \sinh ^2d \exp (i \Delta )+b d \sinh d\big\}.
\end{align}
Then we find the four contributions to the rate of working per wavelength  as
\begin{align}
	&\int_{0}^{2\pi} \left .u_{1,x} \sigma_{1, xy} \right \vert_{y=0} \mathrm{\,d} x \nonumber\\ =& \frac{2\pi\left \{-a^2[\sinh{d}\cosh{d} - d + (d\cosh{d}-\sinh{d})\cos{\Delta}]- abd^2\cos{\phi} + abd\sinh{d}\cos(\Delta+\phi)\right \}}{\sinh^2{d}-d^2}  
	, \\
    &\int_{0}^{2\pi} \left .u_{1,y} \sigma_{1, yy} \right \vert_{y=0} \mathrm{\,d} x \nonumber\\ =& \frac{2\pi\left \{-b^2[\sinh{d}\cosh{d}+d-(d\cosh{d}+\sinh{d})\cos{\Delta}] -abd^2\cos{\phi} - abd\sinh{d}\cos(\Delta-\phi)\right \}}{\sinh^2{d}-d^2}, \\
	&\int_{0}^{2\pi} \left .u_{1,x} \sigma_{1, xy} \right \vert_{y=d} \mathrm{\,d} x \nonumber\\ =& \frac{2\pi\left \{a^2[\sinh{d}\cosh{d} - d + (d\cosh{d}-\sinh{d})\cos{\Delta}] - abd^2\cos{\phi} + abd\sinh{d}\cos(\Delta-\phi)\right \}}{\sinh^2{d}-d^2} , \\
	&\int_{0}^{2\pi} \left .u_{1,y} \sigma_{1, yy} \right \vert_{y=d} \mathrm{\,d} x \nonumber\\ =& \frac{2\pi\left \{b^2[\sinh{d}\cosh{d}+d-(d\cosh{d}+\sinh{d})\cos{\Delta}] - abd^2\cos{\phi} -abd\sinh{d}\cos(\Delta+\phi)\right \}}{\sinh^2{d}-d^2}.
\end{align}
Note that, due to the general lack of  $1\leftrightarrow 2 $ symmetry when the sheets undergo both longitudinal and transverse waving, the contribution to $\dot{W}_2$ from the first sheet is not equal to the contribution from the second. (If one of the waving modes disappears, then both sheets give equal contributions.) 
The total leading-order rate at which work is done by the sheets on the fluid between them per wavelength is finally obtained as
\begin{align}
    \dot{W}_2 =\frac{4\pi}{\sinh^2{d}-d^2} \big\{&a^2[\sinh{d}\cosh{d} - d + (d\cosh{d}-\sinh{d})\cos{\Delta}] \nonumber\\
    &+ b^2[\sinh{d}\cosh{d}+d-(d\cosh{d}+\sinh{d})\cos{\Delta}] \nonumber\\
    &+ 2abd\sinh{d}\sin{\Delta}\sin{\phi}\big\}.\label{eq:W2dot}
\end{align}

\subsection{Minimisation of leading-order rate of working}\label{sec:minimise}

We now consider the minimisation of the leading-order rate of working $\dot{W}_2$ in Eq.~(\ref{eq:W2dot}) with respect to the phase difference $\Delta$ between the waving sheets. Classically, the  rate at which work is done by the sheets  is equal to the rate of  dissipation of mechanical energy in the fluid between the sheets.
Thus, minimising $\dot{W}_2$ is equivalent to finding the optimal phase difference of synchronised sheets, which we denote by $\Delta^*$, leading to minimum dissipation (in our discussion we choose $-\pi\leq\Delta^*<\pi$).

First, we may find by inspection the value of $\Delta^*$ for the two special cases $a=0$ (transverse waving only), which was considered by Taylor, and $b=0$  (longitudinal waving only).
To do this, we note that for $d>0$, the quantities $\sinh d\cosh d \pm d$, $d\cosh d\pm \sinh d$ and $\sinh^2d-d^2$, which   appear in Eq.~(\ref{eq:W2dot}), are all strictly positive. 
Thus, we recover Taylor's result: if the sheets undergo only transverse deformation ($a=0$), then the in-phase waving of the two sheets ($\Delta^*=0$) minimises the rate of working $\dot{W}_2$.
Conversely, if the sheets undergo only longitudinal deformation ($b=0$), then, perhaps surprisingly, the opposite is true. 
The opposite-phase waving  of the two sheets ($\Delta^*=-\pi$) minimises $\dot{W}_2$, while in-phase waving maximises it.
This case will be interpreted further in Sec.~\ref{sec:longitudinal} using a long-wavelength limit of this calculation.

Next we derive the  value of the optimal phase difference $\Delta^*$ for general values  of $a$ and $b$. 
We consider the part of the expression for the rate of working $\dot{W}_2$ that varies with $\Delta$ and  define 
\begin{align}
	S&\equiv\left (a^2-b^2\right )d\cosh d - \left (a^2+b^2\right )\sinh d,\\
	T&\equiv 2abd\sinh d \sin\phi.
\end{align}
These do not depend on the phase difference $\Delta$ between the two sheets, but  are functions of the waving amplitudes $a$ and $b$, the mean sheet separation $d$ and the phase difference between deformation modes $\phi$. 
Then $\dot{W}_2$ as a function of $\Delta$ is minimised when the sum $S\cos\Delta + T\sin\Delta$ is minimised. 
This is equivalent  to minimising $\cos(\Delta-\Theta)$, where
\begin{align}
	\cos\Theta &= \frac{S}{(S^2+T^2)^{\frac{1}{2}}},\\
	\sin\Theta &= \frac{T}{(S^2+T^2)^{\frac{1}{2}}}.
\end{align}
The optimal phase difference $\Delta^*$ is therefore given by
\begin{align}
	\Delta^*=\Theta-\pi. 
\end{align}
Conversely, we see that $\dot{W}_2$ is maximised when $\Delta=\Theta$, which differs from $\Delta^*$ by $\pi$.

\begin{figure}
	\includegraphics[width=\textwidth]{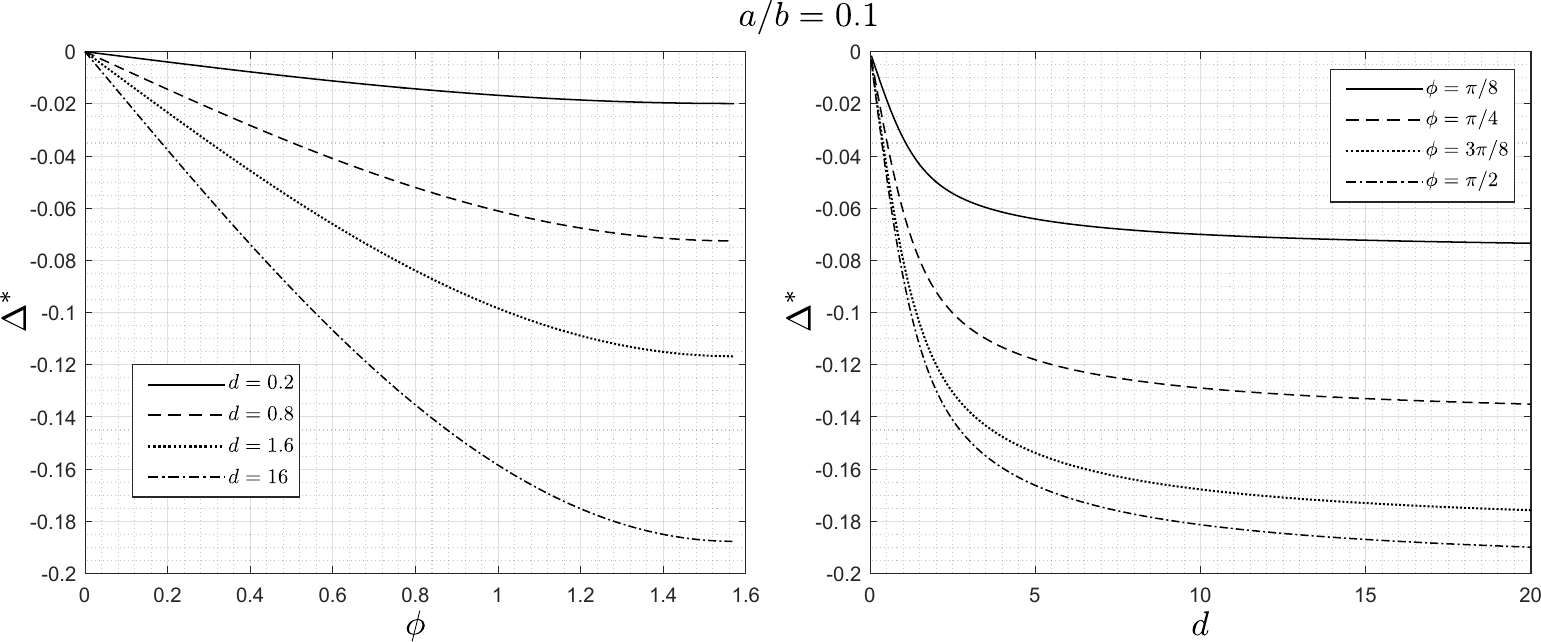}
	\caption{Variation of the optimal phase difference $\Delta^*$ with the phase difference between  the longitudinal and transverse deformation modes  $\phi$ (left panel) and with the mean sheet separation $d$ (right panel). Both panels are  in the case  $a/b=0.1$, i.e.,~mostly transverse waving. 
	The legends indicate the chosen values of $d$ (left) and $\phi$ (right).}
	\label{fig:plot_sheet_Deltadphi_bbig}
\end{figure}
\begin{figure}
	\includegraphics[width=\textwidth]{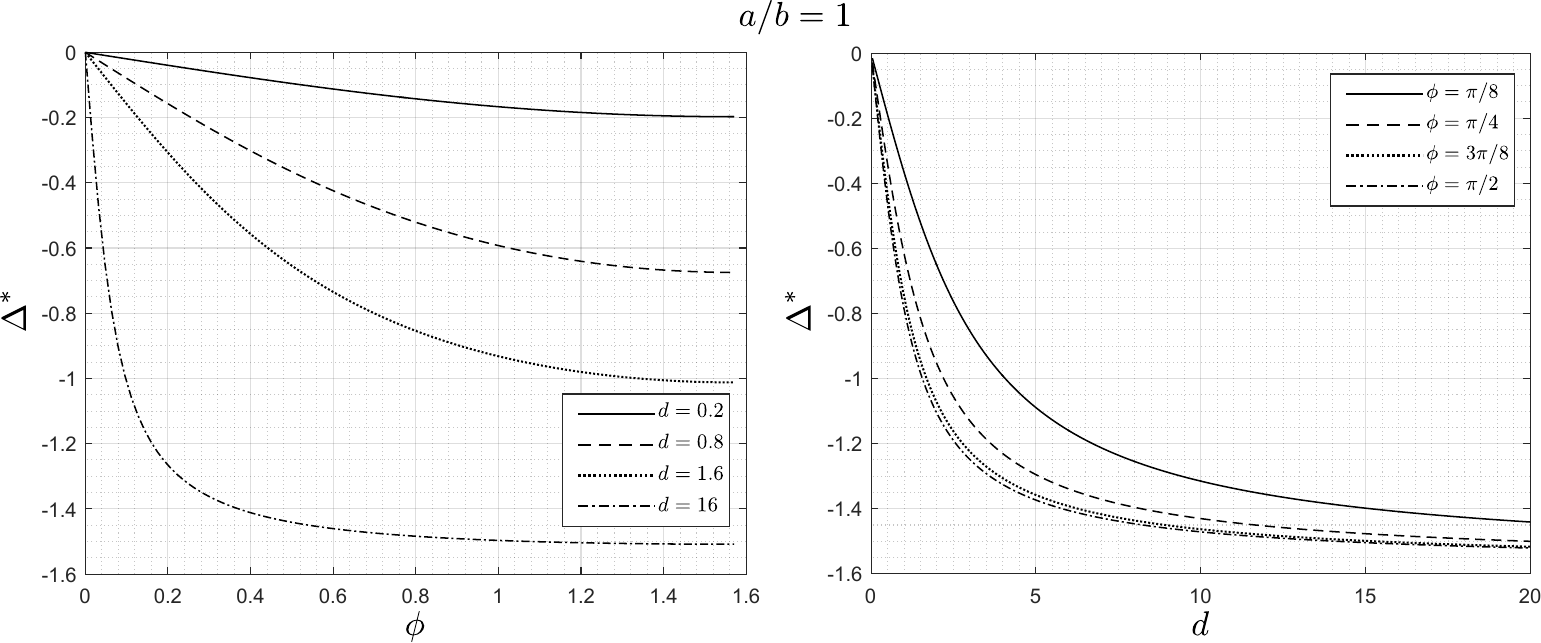}
	\caption{Same as in Fig.~\ref{fig:plot_sheet_Deltadphi_bbig} in the case $a/b=1$, i.e.,~equal longitudinal and transverse waving amplitudes.}
	\label{fig:plot_sheet_Deltadphi_abequal}
\end{figure}
\begin{figure}
	\includegraphics[width=\textwidth]{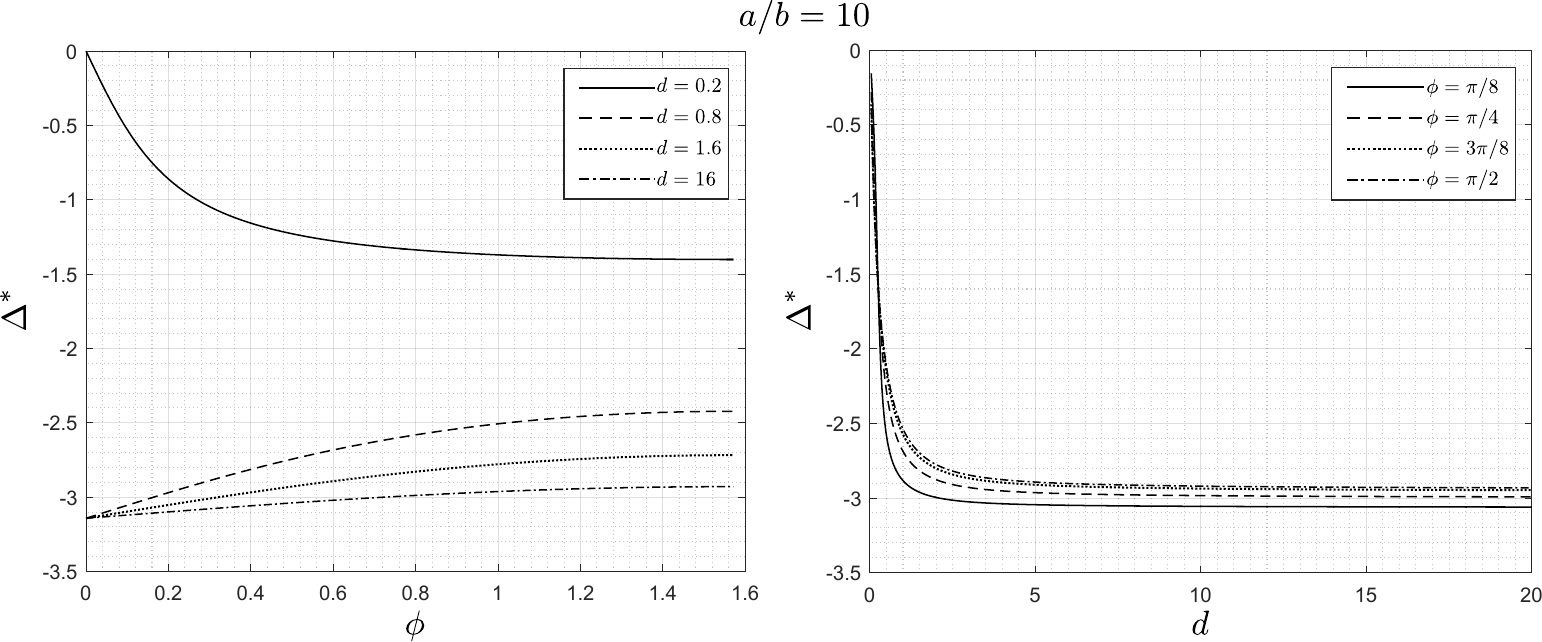}
	\caption{Same as in Fig.~\ref{fig:plot_sheet_Deltadphi_bbig} in the case $a/b=10$, i.e.,~mostly longitudinal waving.}
	\label{fig:plot_sheet_Deltadphi_abig}
\end{figure}

We illustrate the value of the optimal phase difference $\Delta^*$ in  Figs.~\ref{fig:plot_sheet_Deltadphi_bbig},~\ref{fig:plot_sheet_Deltadphi_abequal} and~\ref{fig:plot_sheet_Deltadphi_abig} for three different values of the amplitude ratio $a/b$: $a/b=0.1$ (Fig.~\ref{fig:plot_sheet_Deltadphi_bbig}), 1 (Fig.~\ref{fig:plot_sheet_Deltadphi_abequal}) and 10 (Fig.~\ref{fig:plot_sheet_Deltadphi_abig}).  

On the left panel of each figure, we plot the value of the optimal phase difference $\Delta^*$ as a function of the waving phase $\phi$, for different values of the separation $d$. 
In these plots, we restrict values of $\phi$ so that $0\leq\phi\leq\pi/2$. 
This avoids redundancy, since $T$ is proportional to $\sin\phi$, which has symmetries. Note that if $\phi=0$, then $T=0$. 
This means that $\Delta^*=-\pi$ or $0$ depending on whether $S>0$ or $<0$ respectively.
Then, as the phase $\phi$ increases from 0 to $\pi/2$ with $d$ fixed, $T$ increases, while $S$ remains fixed. Thus, the optimal phase difference $\Delta^*$ also varies monotonically as $\phi$ increases from 0 to $\pi/2$, and whether it increases from $\Delta^*=-\pi$  or decreases from $\Delta^*=0$ depends on the sign of $S$. For example, for $a/b=10$ with $d=0.2$, we have $S<0$, which results in $\Delta^*=0$ for $\phi=0$. This reflects the limit of small $d$, which we   discuss below. However, for $a/b=10$ with the larger values of $d=0.8, 1.6$ and $16$, we have $S>0$, which results in $\Delta^*=-\pi$ for $\phi=0$. This reflects the fact that in these cases the waving is mostly longitudinal; as found earlier, the special case $b=0$ has $\Delta^*=-\pi$.

On the right panel of each figure, we next plot how $\Delta^*$ varies with the mean sheet separation $d$, for selected values of the phase difference $\phi$ between longitudinal and transverse waving modes. In that case, we can examine analytically the two limits $d\to 0$ [which we  compare   with the long-wavelength limit result in Eq.~(\ref{eq:WdotL})]  and  $d\to\infty$. In the limit as $d \to 0$ (i.e.,~sheets very close together), we can expand the expression for $\dot{W}_2$ in powers of $d$ and find 
\begin{align}
\dot{W}_2 = \frac{4 \pi}{5d^3} \left[5a^2 d^2(2+\cos{\Delta}) + 6b^2 \left (5+d^2\right )(1-\cos{\Delta}) + 30abd\sin{\Delta}\sin{\phi} + O\left (d^3\right )\right].\label{eq:W2dotdsmall}
\end{align}
In particular, the leading-order result for $b\neq 0$ is that  in-phase waving always minimises the rate of working. 
This is clearly reflected in the dependence  of $\Delta^*$ on $d$ in Figs.~\ref{fig:plot_sheet_Deltadphi_bbig},~\ref{fig:plot_sheet_Deltadphi_abequal} and~\ref{fig:plot_sheet_Deltadphi_abig} (right panels).
We also see that in general, if the sheets only deform longitudinally ($b=0$), then the rate of working is order $1/d$, but otherwise ($b\neq 0$), the rate of working is order $1/d^3$. Thus, contributions to the rate of working due to longitudinal deformation are  much smaller than those due to transverse deformation.
 
In the opposite limit where $d\to\infty$ (i.e.,~the limit where the sheets are widely separated), we need to distinguish   the cases $a\neq b$ and $a=b$. If $a\neq b$, then $S \sim \left (a^2-b^2\right )d\cosh{d}$ and $T \sim 2abd\sinh{d}\sin\phi$ as $d\to\infty$, so that $\tan\Delta^* \to 2ab\sin\phi/\left(a^2-b^2\right)$ as $d\to\infty$. In the case $a=b$, we instead have $S = -\left (a^2+b^2\right )\sinh{d}$ and $T = 2abd\sinh{d}\sin\phi$. Then we have $\cot\Delta^* =  -\left (a^2+b^2\right )/(2abd\sin\phi) \to 0$ as $d\to\infty$, unless we have $\phi=0$, a case that was treated earlier. These results are reflected in the plots.

We further note that the two plots for $a/b=0.1$ in Fig.~\ref{fig:plot_sheet_Deltadphi_bbig} show that if the waving of the sheets is mainly but not entirely transverse, then $\Delta^*$ is close to zero. This is therefore the relevant limit for inextensible flagella~\cite{taylor51}. 
In other words, the leading-order rate of working is minimised when the sheets wave with a small phase difference between them. 
This result is reminiscent of the small, but nonzero, phase differences   observed in cilia arrays deforming as  metachronal waves.

\subsection{Longitudinal waving: long-wavelength limit and interpretation}\label{sec:longitudinal}
We calculated  above the rate at which work was expended by the swimmers with both longitudinal and transverse   modes in the limit of small-amplitude waving. In particular, we found in Sec.~\ref{sec:minimise} that in the case of only longitudinal waving, the optimal phase difference between the two sheets was $\Delta^*=-\pi$, i.e.,~opposite-phase synchronisation.
This result is the opposite  of  what was obtained by Taylor in the case of transverse waving~\cite{taylor51}. 
To investigate this further, we consider here the long-wavelength   limit, and allow only small-amplitude, longitudinal deformation of the sheets ($b=0$).
In the long-wavelength limit, the mean separation of the swimmers is much smaller than their wavelength, so that the dimensionless mean sheet separation satisfies $d\ll 1$. 
Therefore, we may use the classical lubrication theory of hydrodynamics  to find the rate at which work is done, $\dot{W}_{\text{L}}$, by the swimmers on the fluid between them. 
We will then compare our result to the limit as $d\to 0$ of $\epsilon^2 \dot{W}_2$ previously found for small-amplitude waving, in Eq.~(\ref{eq:W2dotdsmall}), which will allow us to gain physical intuition on the case of  longitudinal waving.

 As in the earlier setup of Sec.~\ref{sec:smallamplsetup}, we impose the dimensionless kinematics of material points on the sheets, given by Eqs.~(\ref{eq:sheetsdimlesskinx1})--(\ref{eq:sheetsdimlesskiny2}) with $b=0$ in the shared swimming frame with coordinates $(x,y)$. 
This results in velocity boundary conditions correct to order $\epsilon$ given by Eqs.~(\ref{eq:sheetsu1x0})--(\ref{eq:sheetsu1yd}).
Since for this lubrication theory calculation we consider only longitudinal deformation of the sheets, we may set the phase difference between the longitudinal and transverse waving modes $\phi$ to zero without loss of generality.
Then the kinematics become
\begin{align}
x_s^{(1)} &= x + \epsilon a \cos(x-t), \\
y_s^{(1)} &= 0, \\
x_s^{(2)} &= x + \epsilon a \cos(x-t+\Delta), \\
y_s^{(2)} &= d .
\end{align}

We now move from the swimming frame to the wave frame, which travels at unit speed in the $x$ direction relative to the swimming frame; that is, the wave frame has coordinates $(\tilde{x},y)$, where $\tilde{x}\equiv x-t$. 
In the wave frame, the velocity at each $\tilde{x}$ on the sheets is constant in time; physically, the wave frame follows the compressions and extensions as they travel along the sheet.

In the long-wavelength limit, the incompressible Stokes equations become the lubrication equations~\cite{leal2007advanced} 
\begin{align}
    \frac{\partial p}{\partial \tilde{x}} &= \frac{\partial^2u_{\tilde{x}}}{\partial y^2}, \\
    \frac{\partial p}{\partial y} &= 0,\\
    \nabla\cdot\mathbf{u}&=0.
\end{align}
The no-slip velocity boundary conditions correct to order $\epsilon$ are given by
\begin{align}
\left .u_{\tilde{x}}\right \vert_{y=0} &= \epsilon a \sin \tilde{x}-1, \\
\left .u_{y}\right \vert_{y=0} &= 0,\\
\left .u_{\tilde{x}}\right \vert_{y=d} &= \epsilon a \sin(\tilde{x}+\Delta)-1, \\
\left .u_{y}\right \vert_{y=d} &= 0.
\end{align}

Since the pressure $p$ is a function of $\tilde{x}$ only, we may directly integrate the $\tilde{x}$ component of the lubrication equations.
Then the horizontal component of the velocity, correct to order $\epsilon$, is given by
\begin{align}
    u_{\tilde{x}} = \frac{1}{2}\frac{\partial p}{\partial \tilde{x}} y(y-d) + \frac{\epsilon a y}{d}  [\sin(\tilde{x}+\Delta)-\sin \tilde{x}] + \epsilon a \sin \tilde{x} -1.\label{eq:longwavelengthux}
\end{align}
This flow is clearly a sum of a pressure-driven flow (quadratic in $y$), a shear flow (linear) and uniform flow (constant). 
The flow rate between the sheets $Q$ is then a function of $\tilde{x}$ given by
\begin{align}
    Q &\equiv \int_0^d u_{\tilde{x}} \mathrm{\,d} y \nonumber\\
    & = -\frac{d^3}{12} \frac{\partial p}{\partial \tilde{x}} + \frac{\epsilon a d}{2} [\sin(\tilde{x}+\Delta) + \sin \tilde{x}] - d.
\end{align}
Using the incompressibility condition and the boundary conditions, we find the derivative of the flow rate 
\begin{align}
\frac{\partial Q}{\partial \tilde{x}} &= \int_0^d \frac{\partial u_{\tilde{x}}}{\partial \tilde{x}} \mathrm{\,d} y \nonumber\\
&= -\int_0^d \frac{\partial u_y}{\partial y} \mathrm{\,d} y \nonumber\\
&= \left .u_{y}\right \vert_{y=0} - \left .u_{y}\right \vert_{y=d}\nonumber\\
&=0.
\end{align}
Therefore, the flow rate $Q$ is constant along the sheet.
Rearranging the expression for $Q$ gives  the pressure gradient
\begin{align}
\frac{\partial p}{\partial \tilde{x}} = \frac{6 }{d^3} \{\epsilon ad[\sin(\tilde{x}+\Delta)+\sin \tilde{x}]-2d-2Q\}.
\end{align}
To eliminate the constant $Q$, we impose the condition that pressure is periodic via 
\begin{align}
\int_0^{2\pi} \frac{\partial p}{\partial \tilde{x}} \mathrm{\,d} \tilde{x} = 0.
\end{align}
This gives $Q=-d$, so that the pressure gradient is obtained explicitly as
\begin{align}
\frac{\partial p}{\partial \tilde{x}} = \frac{6 \epsilon a }{d^2} [\sin(\tilde{x}+\Delta)+\sin \tilde{x}].\label{eq:longwavelengthdpdx}
\end{align}
In the long-wavelength limit, the rate of viscous dissipation in the fluid between the sheets, which is equal to the rate  at which work is done   by the sheets on the fluid between them, $\dot{W}_{\text{L}}$, is given by
\begin{align}
    \dot{W}_{\text{L}} = \int_0^{2\pi} \int_0^d \left(\frac{\partial u_{\tilde{x}}}{\partial y}\right)^2 \mathrm{\,d} y \mathrm{\,d} \tilde{x}.
\end{align}
The shear rate is given by
\begin{align}
    \frac{\partial u_{\tilde{x}}}{\partial y} = \frac{1}{2} \frac{\partial p}{\partial \tilde{x}} (2y-d) + \frac{\epsilon a}{d} [\sin(\tilde{x}+\Delta)-\sin \tilde{x}],
\end{align}
so that the rate of working, correct to order $\epsilon^2$, is finally obtained as 
\begin{align}
    \dot{W}_{\text{L}} &= \frac{4 \pi \epsilon^2 a^2 }{d}(2+\cos{\Delta}).\label{eq:WdotL}
\end{align}

Earlier, we calculated the order $\epsilon^2$ rate of working for small-amplitude waving, and found the expression in Eq.~(\ref{eq:W2dotdsmall}) for $\dot{W}_2$, valid for small mean sheet separation $d$.
The expression for $\dot{W}_{\text{L}}$ in Eq.~(\ref{eq:WdotL}) agrees with $\epsilon^2\dot{W}_2$ from Eq.~(\ref{eq:W2dotdsmall}) for $b=0$ (longitudinal waving only) to leading order. 
Therefore, in the lubrication limit, we recover the result for $d \ll 1$ and $b=0$ that opposite-phase waving minimises the rate at which work is done by the sheets. 

\begin{figure}
\includegraphics[width=0.5\textwidth]{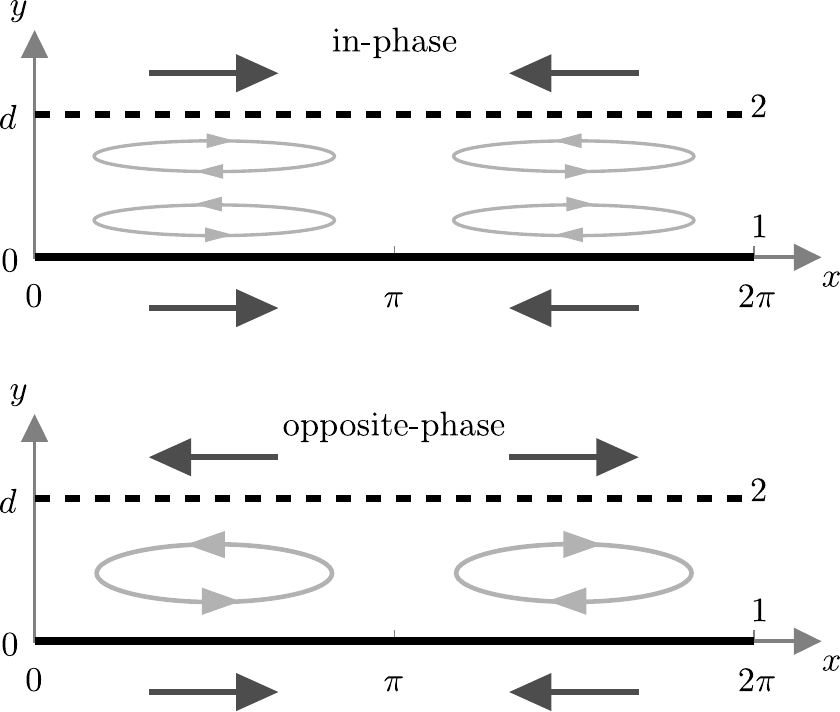}
\caption{Sketch of flow fields for in-phase (top) and opposite-phase (bottom) small-amplitude, longitudinal deformation of sheets in the swimming frame $(x,y)$. 
Arrows indicate direction of motion of the sheets (dark arrows) and the fluid (light arrows).
In-phase waving of the sheets induces backflows halfway between the swimmers and increases the total rate of viscous dissipation.}
\label{fig:sketch_longwavelength_inoppositephase}
\end{figure}

To help explain intuitively why the value of $\Delta^*=-\pi$ leads to minimum energy dissipation for synchronised longitudinal waving, we sketch in Fig.~\ref{fig:sketch_longwavelength_inoppositephase} the flow fields for in-phase (top) and opposite-phase (bottom) waving of the sheets in the swimming frame, so that we may see the order $\epsilon$ flow clearly. From the results above, we note that the pressure gradient given by Eq.~(\ref{eq:longwavelengthdpdx}) may be written as
\begin{align}
\frac{\partial p}{\partial \tilde{x}} = \frac{12 \epsilon a }{d^2}\cos \left (\frac{\Delta}{2}\right )\sin\left (\tilde{x}+\frac{\Delta}{2}\right ).
\end{align}
The periodic pressure gradient  has an amplitude  proportional to $\left \lvert\cos(\Delta/2) \right \rvert$.  In particular, the pressure gradient is zero at the optimal phase difference $\Delta^*=-\pi$ between the sheets, for which  there is no quadratic part of the flow in Eq.~(\ref{eq:longwavelengthux}). In that  case, in the swimming frame, the fluid circulates in cells of width $\pi$ and height $d$ between the two sheets in a linear, shearlike manner. 
If instead the sheets wave in phase ($\Delta=0$), then in the swimming frame, backflows 
 are induced (relative to the waving of the swimmers at each $x$) halfway between the sheets due to incompressibility. 
The fluid circulates in smaller cells  of width $\pi$ and smaller height $d/2$ between the two swimmers, again in the swimming frame.
These smaller recirculation regions are accompanied by the maximum possible amplitude of pressure gradient and the maximum rate of working as $\Delta$ is varied.

\section{Model for interacting cilia:  Two spheres in elliptical orbits}\label{sec:spheres}

After considering a model for two interacting flexible flagella in the previous section, we 
 now investigate the energetics of synchronised cilia  interacting through the fluid.
 
 Following a classical modelling approach, we represent the   anchored cilia as     rigid spheres immersed  in a fluid and orbiting periodically above a no-slip surface. We consider two such identical spheres that interact hydrodynamically in the far field.
The flow outside the spheres is governed by the incompressible Stokes equations for a fluid of viscosity $\mu$, Eqs.~(\ref{eq:Stokes})--(\ref{eq:incompressible}), and the far-field flow generated by the motion of a sphere will be approximated as that due to a point force (stokeslet)~\cite{kimbook}.
Each sphere undergoes periodic motion in an elliptical orbit with a constant imposed phase difference $\Delta$.
An elliptical orbit is a minimal model for the two-stroke motion of a flexible cilium, capturing the essential time-irreversibility that allows it to generate net forces and flows. 
Here, similarly to the previous section, we impose the kinematics of each sphere and calculate their time-averaged rate of working  as a function of the phase difference $\Delta$. 
We will consider two different cases. 
In Sec.~\ref{sec:parallelcircular}, the orbits lie in a plane parallel to the no-slip surface and are circular, similar to the setup in Ref.~\cite{niedermayer08}.  
Next, in Sec.~\ref{sec:perpendicularelliptical}, each sphere's orbit lies in a plane perpendicular to the no-slip surface and is elliptical, a motion akin to  that of the centres of mass of flexible cilia during their effective and recovery strokes.

\subsection{Circular orbits parallel to no-slip surface}\label{sec:parallelcircular}

\subsubsection{Setup}

\begin{figure}
	\includegraphics[width=0.6\textwidth]{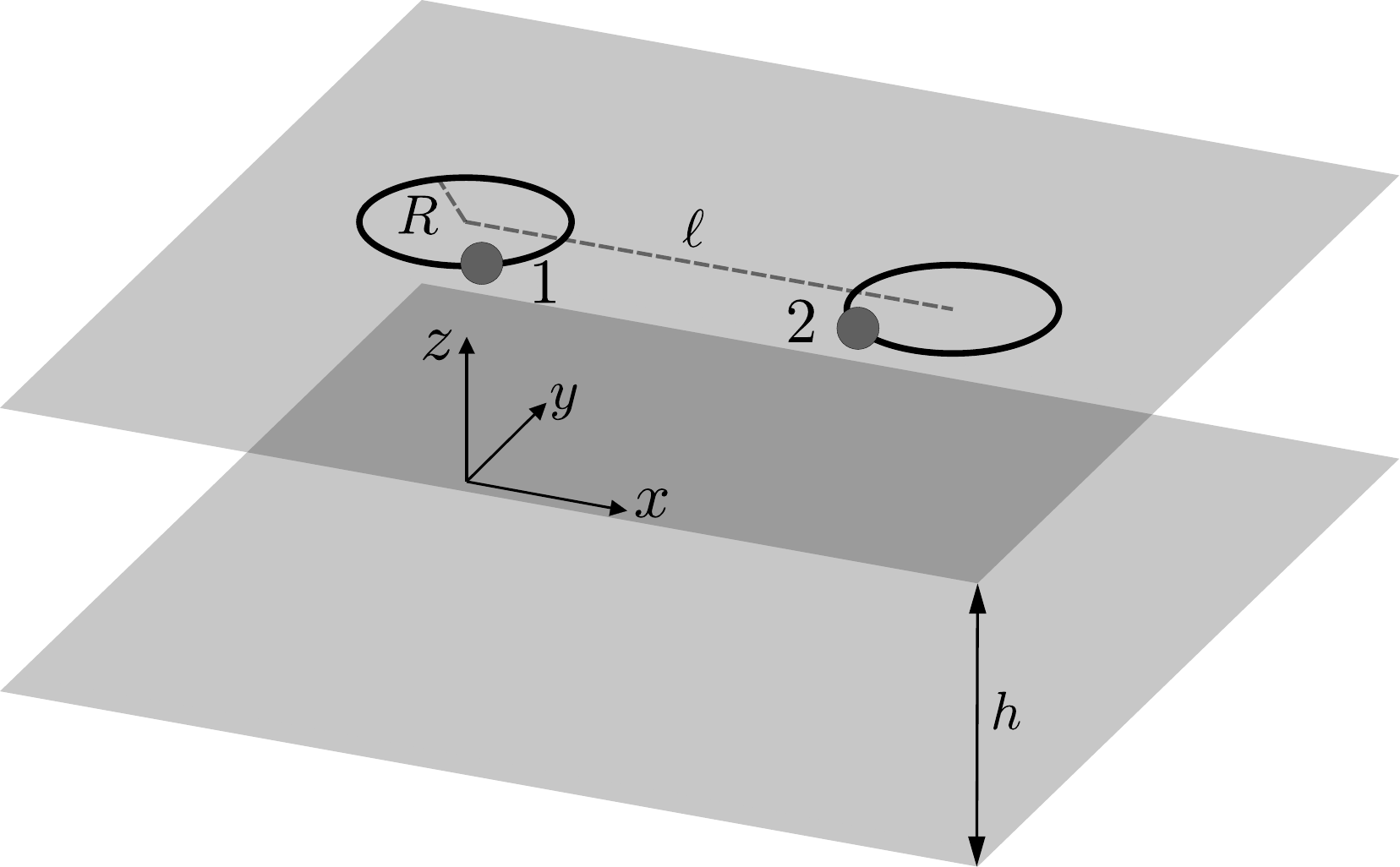}
	\caption{Two identical spheres  interacting hydrodynamically 	are models for two anchored cilia. 
	Each sphere is of radius $a$ and moves in a circular orbit of radius $R$. 
	The orbits lie in a plane a distance $h$ above a parallel no-slip surface  and their centres are separated by distance $\ell$.}
	\label{fig:diagram_sphere_circle}
\end{figure}
\begin{figure}
	\includegraphics[width=0.5\textwidth]{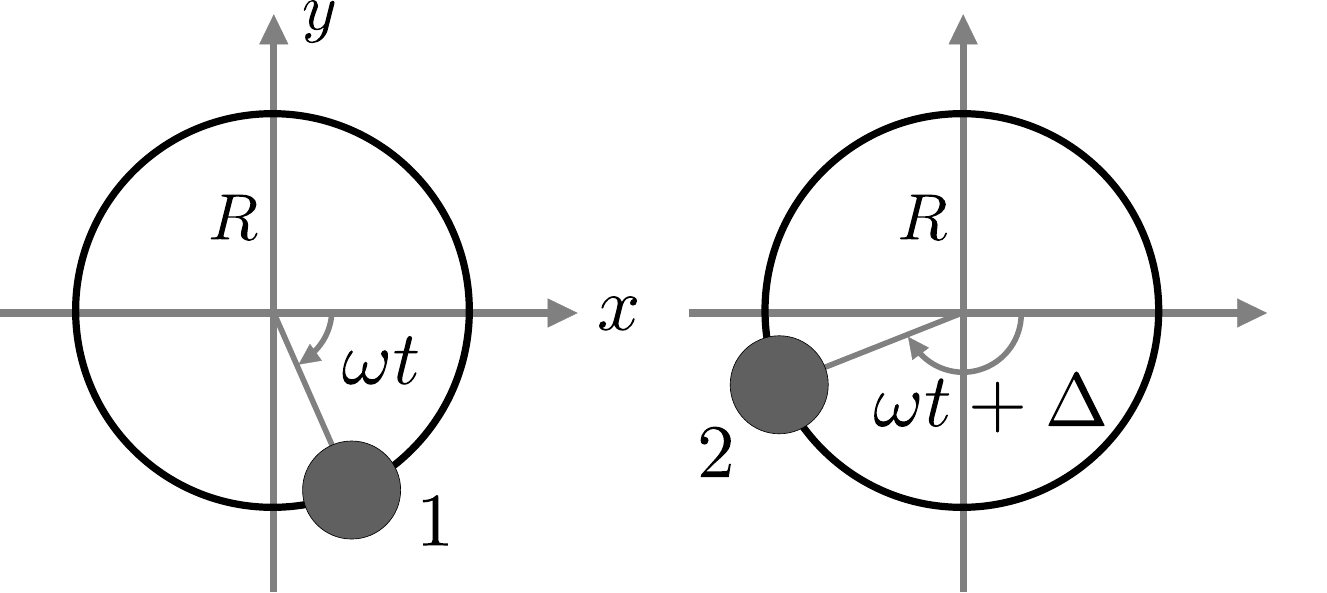}
	\caption{View from above of each orbit in Fig.~\ref{fig:diagram_sphere_circle}.}
	\label{fig:diagram_sphere_circle_orbit}
\end{figure}

The setup is illustrated in Fig.~\ref{fig:diagram_sphere_circle}, with views of each orbit from above shown in Fig.~\ref{fig:diagram_sphere_circle_orbit}.  Each sphere, of radius $a$, moves at constant angular speed $\omega$ in a circular orbit of radius $R$, clockwise when viewed from above, with a constant imposed phase difference of $\Delta$. Each orbit lies in the plane a distance $h$ above the no-slip surface.  We use Cartesian coordinates $(x,y,z)$ with corresponding unit vectors $\mathbf{e}_x$, $\mathbf{e}_y$, and $\mathbf{e}_z$, where the plane $z=0$ is the no-slip surface, and the centres of the orbits of the spheres are separated by a distance $\ell$ in the $x$ direction. Thus, we may choose the centres of the orbits of the first and second spheres to have coordinates $(0,0,h)$ and $(\ell,0,h)$ respectively. The spheres are far from the no-slip surface ($a\ll h$).  In order to derive our results analytically, and as done in other work, we  let the separation of orbit centres $\ell$ be much larger than all other imposed length scales in the setup ($a,h,R\ll\ell$). This means that the spheres are  widely separated at all times and interact hydrodynamically only in the far field. We proceed to calculate the rate at which work is done by each sphere averaged over one period, including the first correction due to interaction between the two spheres, which depends on the phase difference $\Delta$.  

We impose the kinematics as follows. 
The position of the first sphere  $\mathbf{r}_1$ is a function of time $t$ given by
\begin{align}
    \mathbf{r}_1 = R(\cos\omega t\,\mathbf{e}_x -\sin\omega t\,\mathbf{e}_y) + h\mathbf{e}_z,
\end{align}
so that its velocity  is  
\begin{align}
    \mathbf{U}_1 &= \frac{\mathrm{\, d} \mathbf{r}_1}{\mathrm{\, d} t}\nonumber\\ 
    &= -\omega R(\sin\omega t\,\mathbf{e}_x +\cos\omega t\,\mathbf{e}_y).
\end{align}
The position of the second sphere $\mathbf{r}_2$ is given by 
\begin{align}
    \mathbf{r}_2 &= R[\cos(\omega t + \Delta)\mathbf{e}_x -\sin(\omega t + \Delta)\mathbf{e}_y] + \ell\mathbf{e}_x+h\mathbf{e}_z,
\end{align}
and its velocity is
\begin{align}
    \mathbf{U}_2 &= \frac{\mathrm{\, d} \mathbf{r}_2}{\mathrm{\, d} t}\nonumber\\ 
    &=  -\omega R[\sin(\omega t + \Delta)\mathbf{e}_x +\cos(\omega t + \Delta)\mathbf{e}_y].
\end{align}

\subsubsection{Time-averaged rate of working}

From the setup above, we can calculate the time-averaged rate at which work is done by the first sphere in the far-field limit, by using Fax\'{e}n's first law.
We denote by $\mathbf{u}_{2\to 1}$ the flow induced by the motion of the second sphere at the position of the first sphere and let $\mathbf{r}_{2\to 1}$ be the position of the first sphere relative to the second (i.e.,~$\mathbf{r}_{2\to 1}\equiv \mathbf{r}_1-\mathbf{r}_2$). 
We also denote by $\mathbf{F}_i$ the force exerted by the $i$th sphere on the fluid.
The flow $\mathbf{u}_{2\to 1}$ may be approximated as that due to a point force $\mathbf{F}_2$ above a no-slip surface, since the spheres are widely separated.
Using the classical image system for flow singularities above rigid walls~\cite{blake1974fundamental}, and since the spheres are the same distance $h$ above the no-slip surface, the flow $\mathbf{u}_{2\to 1}$ is given to leading order by
\begin{align}
	\mathbf{u}_{2\to 1}\simeq\frac{3h^2}{2\pi\mu}\frac{(\mathbf{F}_2\cdot\mathbf{r}_{2\to 1})\mathbf{r}_{2\to 1}}{\left \lvert \mathbf{r}_{2\to 1}\right \rvert^5}.\label{eq:u2to1Blake}
\end{align}
Approximating the second sphere generating the force  $\mathbf{F}_2$ above
 as moving in an unbounded  fluid otherwise at rest, that force   is given by Stokes's law as
\begin{align}
	\mathbf{F}_2\simeq 6 \pi \mu a \mathbf{U}_2.\label{eq:F2U2}
\end{align}
In other words, we neglect  at leading order the impact of the flow due to the first sphere at the position of the second on the hydrodynamic interactions, since this decays in the limit as the spheres become infinitely separated. We also neglect order $a/h$ corrections to the viscous resistance coefficient for the sphere ($\simeq 6\pi\mu a$), since the sphere is assumed to be far from the no-slip surface. To leading order we have approximately $
	\mathbf{r}_{2\to 1}\simeq -\ell\mathbf{e}_x,
$ since the radius of orbit is small, $R\ll \ell$.
Then the flow $\mathbf{u}_{2\to 1}$ is given to leading order by
\begin{align}
\mathbf{u}_{2 \to 1} \simeq - \frac{9 a h^2}{\ell^3} \omega R \sin(\omega t + \Delta) \mathbf{e}_x.
\end{align}

This then allows us to obtain  the leading-order correction to the Stokes's law value for  the force $\mathbf{F}_1$ exerted by the first sphere on the fluid. 
This perturbation to the value given by Stokes's law is due to the flow created  by second sphere in the far field. 
By Fax\'{e}n's first law, the force $\mathbf{F}_1$ exerted by the first sphere on the fluid  is approximately given by  
\begin{align}
    \mathbf{F}_1 &\simeq 6 \pi \mu a (\mathbf{U}_1 - \mathbf{u}_{2 \to 1})\nonumber\\
    &\simeq 6\pi\mu a\omega R\left(-(\sin\omega t\,\mathbf{e}_x +\cos\omega t\,\mathbf{e}_y)+ \frac{9 a h^2}{\ell^3} \sin(\omega t + \Delta) \mathbf{e}_x\right ).\label{eq:Faxen}
\end{align}
This gives the instantaneous rate of working
\begin{align}
	\mathbf{F}_1\cdot\mathbf{U}_1\simeq 6 \pi \mu a\omega^2 R^2 \left (1 - \frac{9 a h^2}{\ell^3}  \sin\omega t\sin(\omega t + \Delta)\right ).
\end{align}

Then the rate at which work is done by the first sphere averaged over a period is obtained as
\begin{align}\label{eq:98_final}
    &\frac{\omega}{2\pi} \int_0^{{2\pi}/{\omega}} \mathbf{F}_1 \cdot \mathbf{U}_1 \mathrm{\, d} t \simeq 6 \pi \mu a \omega^2 R^2 \left(1 - \frac{9ah^2}{2\ell^3} \cos{\Delta}\right).
\end{align}
To find the corresponding result for the second sphere, we replace the phase difference $\Delta$ with $-\Delta$ and find that the rate of  working is unchanged;  the total rate at which energy is dissipated in the fluid is thus twice the result from Eq.~\eqref{eq:98_final}.  
Consequently,  the in-phase synchronised  motion of the spheres always minimises the average rate of dissipation of energy in the fluid. We will compare these results to past predictions for  the dynamic synchronisation in Sec.~\ref{sec:discussion}.

\subsection{Elliptical orbits perpendicular to no-slip surface}\label{sec:perpendicularelliptical}

\subsubsection{Setup}
\begin{figure}
	\includegraphics[width=0.8\textwidth]{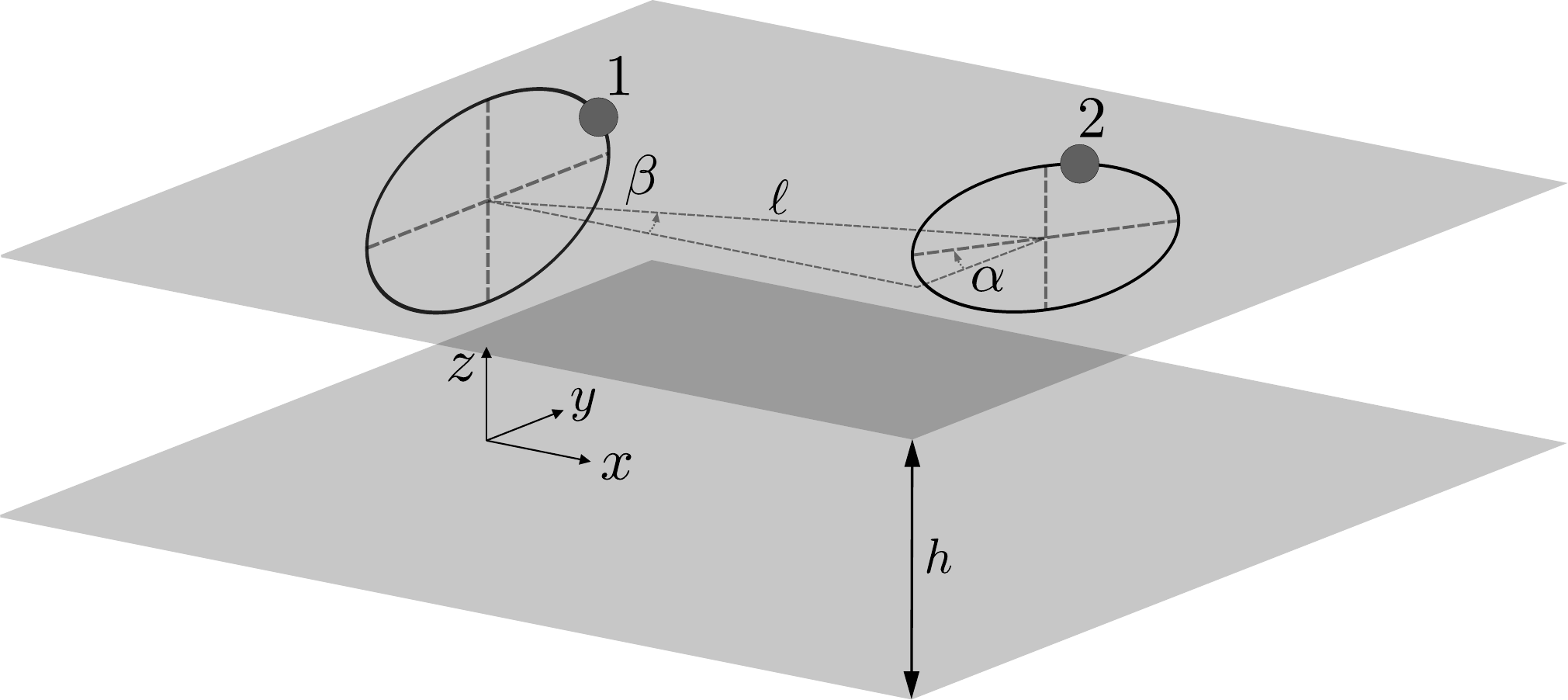}
	\caption{Two identical spheres of radius $a$ interacting hydrodynamically as models for two anchored cilia. 
	Each orbit is elliptical and has centre at height $h$ above the no-slip surface $z=0$. 
	Each orbit lies in a plane perpendicular to the no-slip surface.  
	The relative angle  between the planes of the two orbits is $\alpha$ (see top view in Fig.~\ref{fig:diagram_sphere_ellipse_top}). 
	The centre of the second sphere's orbit is displaced from that of the first sphere's orbit by the vector $\ell\cos\beta\mathbf{e}_x + \ell\sin\beta\mathbf{e}_y$.}
	\label{fig:diagram_sphere_ellipse}
\end{figure}
\begin{figure}
	\includegraphics[width=0.5\textwidth]{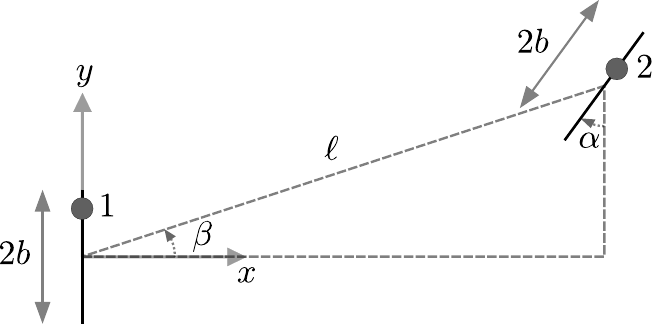}
	\caption{View from above of setup in Fig.~\ref{fig:diagram_sphere_ellipse}.
	Each elliptical orbit has one semi-axis of length b lying in the plane $z=h$.}
	\label{fig:diagram_sphere_ellipse_top}
\end{figure}

In Sec.~\ref{sec:parallelcircular}, we considered a model in which the orbits of the spheres are circular and lie in a plane parallel to a no-slip surface. This allowed us to consider the energetic properties of the setup addressed in the classical work of Ref.~\cite{niedermayer08}. 
In this section, we now allow the orbits to be elliptical and oriented  perpendicular to the no-slip surface. 
Each of the two identical rigid spheres of radius $a$ may be viewed as representing the centre of mass of a flexible cilium~\cite{brumley2014flagellar}. 
The elliptical orbit is therefore an approximation to the cilium's two-stroke periodic motion, with an effective stroke where the cilium is further from the surface and a recovery stroke where the cilium is closer to the surface.
We again will calculate the rate at which work is done by each sphere as a function of the phase difference $\Delta$ between their orbiting motions.

The setup is illustrated in Fig.~\ref{fig:diagram_sphere_ellipse}, with a view from above in Fig.~\ref{fig:diagram_sphere_ellipse_top}. 
Each sphere moves in an elliptical orbit with semi-axes $b$ and $c$ lying in a plane perpendicular to the no-slip surface. 
The centres of the orbits are separated by a distance $\ell$ and lie a distance $h$ above the no-slip plane, while the relative angle between the two orbital planes is $\alpha$.
The imposed motion of each sphere is identical but with a constant imposed phase difference of $\Delta$.
The period of the motion is $2\pi/\omega$, but importantly we note that the frequency $\omega$ is not the angular velocity of the spheres, since the orbits are not circular in general. 
In other words, unless the orbit is circular, the motion does not occur at constant speed. 

We again use Cartesian coordinates $(x,y,z)$ with corresponding unit vectors $\mathbf{e}_x$, $\mathbf{e}_y$ and $\mathbf{e}_z$, where the no-slip surface is given by $z=0$.
We choose the orbit of the first sphere to have centre with coordinates $(0,0,h)$, and   its semi-axis of length $b$ aligned with the $y$ axis when viewed from above as in Fig.~\ref{fig:diagram_sphere_ellipse_top}.
The semi-axes of the orbits have  length $c$ parallel to the $z$ axis.
We let the centre of the second sphere's orbit  have coordinates $(\ell\cos\beta,\ell\sin\beta,h)$.
The second sphere's orbit, when viewed from above, makes an angle $\alpha$ with the $y$ axis.
This setup breaks the symmetry between the two spheres.
The spheres are at all times far from the no-slip surface ($a,c\ll h$) and widely separated ($a,b\ll\ell$).
The separation $\ell$ of the centres of the orbits is the largest imposed length scale ($\ell\gg h$), allowing us to treat the
hydrodynamic interactions between the spheres   in the far field.

We impose the kinematics as follows. The position of the first sphere $\mathbf{r}_1$ is given by 
\begin{align}
\mathbf{r}_1=b\cos\omega t \, \mathbf{e}_{y} + c \sin\omega t \,\mathbf{e}_{z} + h\mathbf{e}_z,
\end{align}
so that its velocity  $\mathbf{U}_1$ is  
\begin{align}
\mathbf{U}_1 &= \frac{\mathrm{\, d} \mathbf{r}_1}{\mathrm{\, d} t}\nonumber\\
&= \omega(-b\sin\omega t \, \mathbf{e}_{y} +  c \cos\omega t \,\mathbf{e}_{z}).
\end{align}
Similarly, the  second sphere has position $\mathbf{r}_2$ given by 
\begin{align}
\mathbf{r}_2=b\cos(\omega t + \Delta) (\sin\alpha\mathbf{e}_{x}+\cos\alpha\mathbf{e}_y) + c \sin(\omega t + \Delta) \mathbf{e}_{z} + \ell\cos\beta\mathbf{e}_x+\ell\sin\beta\mathbf{e}_y+h\mathbf{e}_z,
\end{align}
with velocity  $\mathbf{U}_2$   given by
\begin{align}
\mathbf{U}_2 &= \frac{\mathrm{\, d} \mathbf{r}_2}{\mathrm{\, d} t}\nonumber\\
&= \omega[- b\sin(\omega t + \Delta)(\sin\alpha\mathbf{e}_{x}+\cos\alpha\mathbf{e}_y)  +  c \cos(\omega t+\Delta) \mathbf{e}_{z}].
\end{align}

\subsubsection{Time-averaged rate of working}

As in Sec.~\ref{sec:parallelcircular}, we use Fax\'{e}n's first law to find the time-averaged rate at which work is done by the first sphere in the far-field limit. 
We also retain the notation $\mathbf{u}_{2\to 1}$ for the flow induced by the motion of the second sphere at the position of the first sphere and  $\mathbf{r}_{2\to 1}\equiv \mathbf{r}_1-\mathbf{r}_2$ for the relative position vector. As above we denote by $\mathbf{F}_i$   the force exerted by the $i$th sphere on the fluid. 
Since we are in the far field, we may approximate the flow $\mathbf{u}_{2\to 1}$ as that due to a point force $\mathbf{F}_2$. Furthermore, we may  use the leading-order expression for $\mathbf{u}_{2\to 1}$ in Eq.~(\ref{eq:u2to1Blake}), since the orbits are small compared with the distance $h$ to the no-slip plane ($a,c\ll h$). As in Eq.~(\ref{eq:F2U2}), the force $\mathbf{F}_2$ is then obtained   to leading order as 
\begin{align}
\mathbf{F}_2\simeq 6 \pi \mu a \mathbf{U}_2,
\end{align}
since the spheres are widely separated and far from the no-slip surface.
This time, however, we have 
\begin{align}
	\mathbf{r}_{2\to 1} \simeq -\ell(\cos\beta\mathbf{e}_x+\sin\beta\mathbf{e}_y),
\end{align}
which gives the leading-order flow 
\begin{align}
\mathbf{u}_{2 \to 1} &\simeq \frac{9ah^2 }{ \ell^3} \left [\left(U_{2,x} \cos^2\beta + U_{2,y} \cos\beta\sin\beta\right) \mathbf{e}_x + \left(U_{2,x} \cos\beta\sin\beta + U_{2,y} \sin^2\beta\right) \mathbf{e}_y  \right ]\nonumber\\
&= -\frac{9 a b \omega h^2 \sin(\omega t + \Delta)\sin(\beta+\alpha)}{\ell^3}(\cos\beta \mathbf{e}_x + \sin\beta\mathbf{e}_y ).
\end{align}

As above, we can now   find the leading-order correction to the Stokes's law value 
for the force $\mathbf{F}_1$ exerted by the first sphere, due to the motion of the second sphere. 
Substituting quantities into Fax\'{e}n's first law gives
\begin{align}
\mathbf{F}_1 &\simeq 6 \pi \mu a (\mathbf{U}_1 - \mathbf{u}_{2 \to 1})\nonumber\\
&\simeq 6\pi\mu a\omega\left (-b\sin\omega t \, \mathbf{e}_y + c\cos\omega t \,\mathbf{e}_z + \frac{9 a b  h^2 \sin(\omega t + \Delta)\sin(\beta+\alpha)}{\ell^3}(\cos\beta \mathbf{e}_x + \sin\beta\mathbf{e}_y ) \right )
\end{align}
and
\begin{align}
&\mathbf{F}_1 \cdot \mathbf{U}_1 \simeq \nonumber \\
& 6 \pi\mu a \omega^2 \left( b^2  \sin^2\omega t + c^2  \cos^2\omega t - \frac{9 a b^2 h^2  \sin\omega t \sin(\omega t + \Delta)\sin\beta\sin(\beta+\alpha)}{\ell^3}    \right ).
\end{align}

\begin{figure}
	\includegraphics[width=0.6\textwidth]{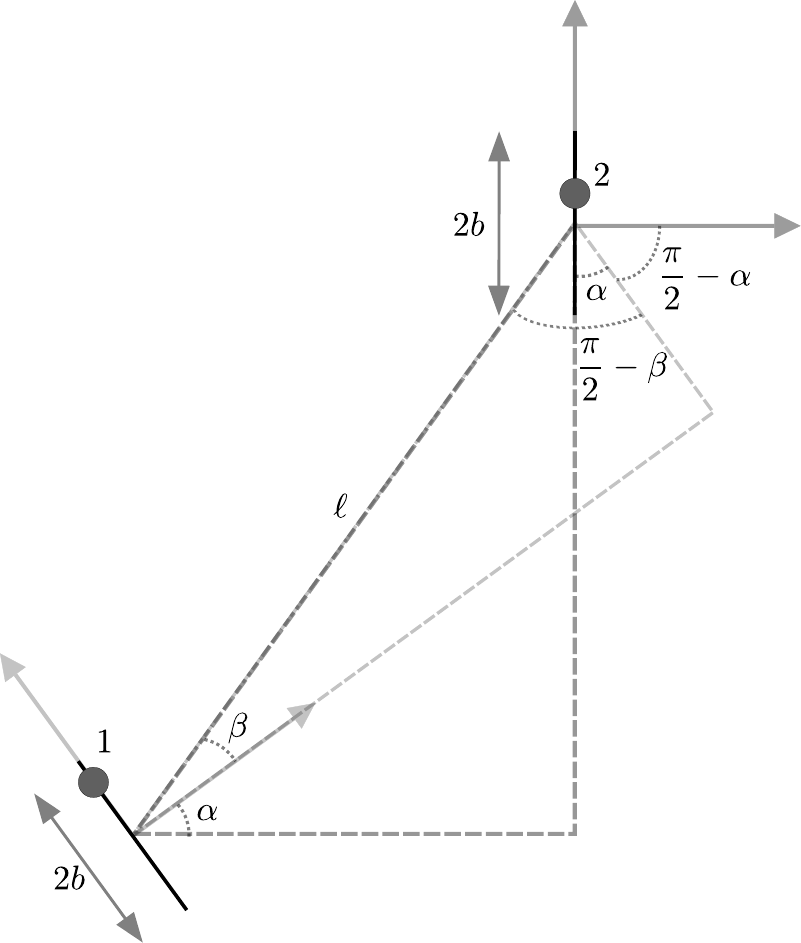}
	\caption{Diagram illustrating the required changes in angles in the setup of Fig.~\ref{fig:diagram_sphere_ellipse_top} to derive the rate at which work is done by the second sphere from that by the first sphere. We replace the angle $\beta$ with $-[(\pi/2-\alpha) + (\pi/2-\beta)]=\alpha+\beta-\pi$, the relative angle $\alpha$ with $-\alpha$ and the phase difference $\Delta$ with $-\Delta$.}
	\label{fig:diagram_sphere_ellipse2angles}
\end{figure}

Then the rate of  work done by the first sphere averaged over a period is obtained as 
\begin{align}
&\frac{\omega}{2\pi} \int_0^{2\pi/\omega} \mathbf{F}_1 \cdot \mathbf{U}_1 \mathrm{\, d} t  \simeq 3\pi\mu a \omega^2\left (b^2 + c^2  - \frac{9 a b^2  h^2 \sin\beta\sin(\beta+\alpha)  \cos{\Delta} }{\ell^3}\right ).
\end{align}
To find the corresponding result for the second sphere, we may use a symmetry argument  illustrated  in 
 Fig.~\ref{fig:diagram_sphere_ellipse2angles} (which was not evident a priori since the geometrical setup is not symmetric).  If we 
 replace the angle $\beta$ with $\alpha+\beta-\pi$, the relative angle $\alpha$ with $-\alpha$ and the phase difference $\Delta$ with $-\Delta$, we see that the setup has been switched $1\leftrightarrow 2$. 
 As a result, we obtain that the  rate of working of the second sphere is equal to the rate of working of the first sphere at this order.

\begin{figure}
	\includegraphics[width=0.5\textwidth]{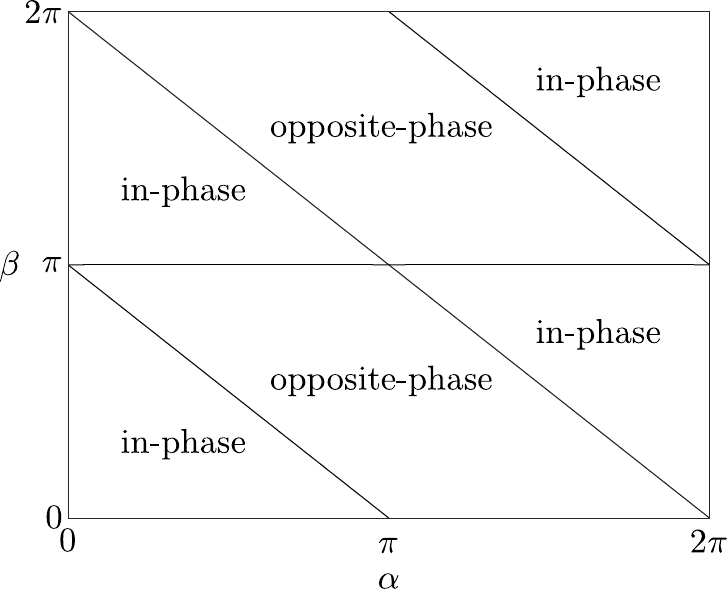}
	\caption{Regions in the $(\alpha,\beta)$ plane for the setup in Fig.~\ref{fig:diagram_sphere_ellipse}  where in-phase or opposite-phase motion of spheres minimises time-averaged rate of working. The relative angle between the planes of the orbits is $\alpha$. The angle $\beta$ indicates the relative position of the centres of the  orbits.}
	\label{fig:plot_sphere_ellipse_optimal}
\end{figure}

With this, we can now consider the optimisation of the viscous dissipation with respect to the phase difference $\Delta$ between the synchronised  motion of the two spheres.
The sign of the coefficient of $\cos\Delta$ determines whether rate of working is minimised by in-phase or opposite-phase motion.
This depends on the quantity $\sin\beta\sin(\beta+\alpha)$, which is unchanged by replacing $\beta$ with $\pi+\beta$ and whose sign is changed by replacing $\alpha$ with $\pi+\alpha$. 
If in-phase motion minimises the rate of working, then opposite-phase motion maximises it, and vice versa.
We see that if $\sin\beta\sin(\beta+\alpha)$ is positive, then in-phase motion of the spheres minimises the rate of working; that is, motion where the two spheres are at the same height at all times leads to minimum power dissipated in the fluid.
This includes the case of precisely aligned elliptical orbits ($\alpha=0$).
On the other hand, if $\sin\beta\sin(\beta+\alpha)$ is negative, then  opposite-phase motion of the spheres is the optimal motion. 
These results are  illustrated in a plot of the $(\alpha,\beta)$ plane in Fig.~\ref{fig:plot_sphere_ellipse_optimal}.
We see that if the two orbital planes are almost but not precisely aligned ($\alpha$ close to 0 or $2\pi$), then in-phase motion minimises energy dissipation for most values of $\beta$, while opposite-phase motion leads to the minimum for a narrow range of $\beta$ near $\pi$.  The same result  (either in-phase or opposite-phase motion is optimal, depending on geometry) was obtained for the planar beating of three-dimensional flagella~\cite{mettot2011energetics}. In  Sec.~\ref{sec:discussion},  we will compare these results to related studies  showing that both in-phase and  opposite-phase beating can  also arise dynamically, depending on the relative conformation of model  cilia~\cite{vilfan06}.

\section{Model for interacting nodal cilia: Two whirling rods}\label{sec:rods}

In this final model, we consider nodal cilia,  which, instead of beating with   effective and recovery strokes of different shapes, rotate rigidly along conical surfaces. We use the whirling rod model for nodal cilia proposed in past studies~\cite{smith2008fluid,smith2011mathematical} to compute the impact of phase difference on the rate of  working of two such interacting, synchronised  cilia.

\subsection{Setup}

\begin{figure}
	\includegraphics[width=0.7\textwidth]{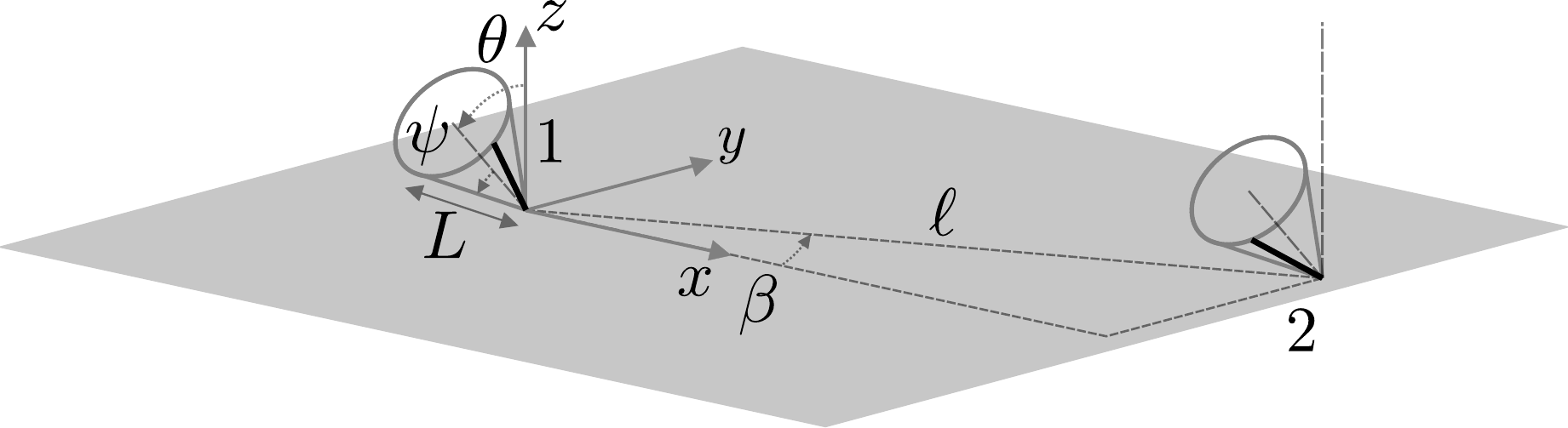}
	\caption{Two identical rigid rods of length $L$ as models for two nodal cilia interacting hydrodynamically. 
	Each rod rotates clockwise when viewed from above, sweeping out a cone with tip fixed on a no-slip surface.
	The semi-angle of each cone is $\psi$  and each cone has a posterior tilt of $\theta$ about the $x$-axis. 
	The second cone is displaced from the first cone by $\ell\cos\beta\mathbf{e}_x + \ell\sin\beta\mathbf{e}_y$.}
	\label{fig:diagram_nodal}
\end{figure}

\begin{figure}
	\includegraphics[width=0.5\textwidth]{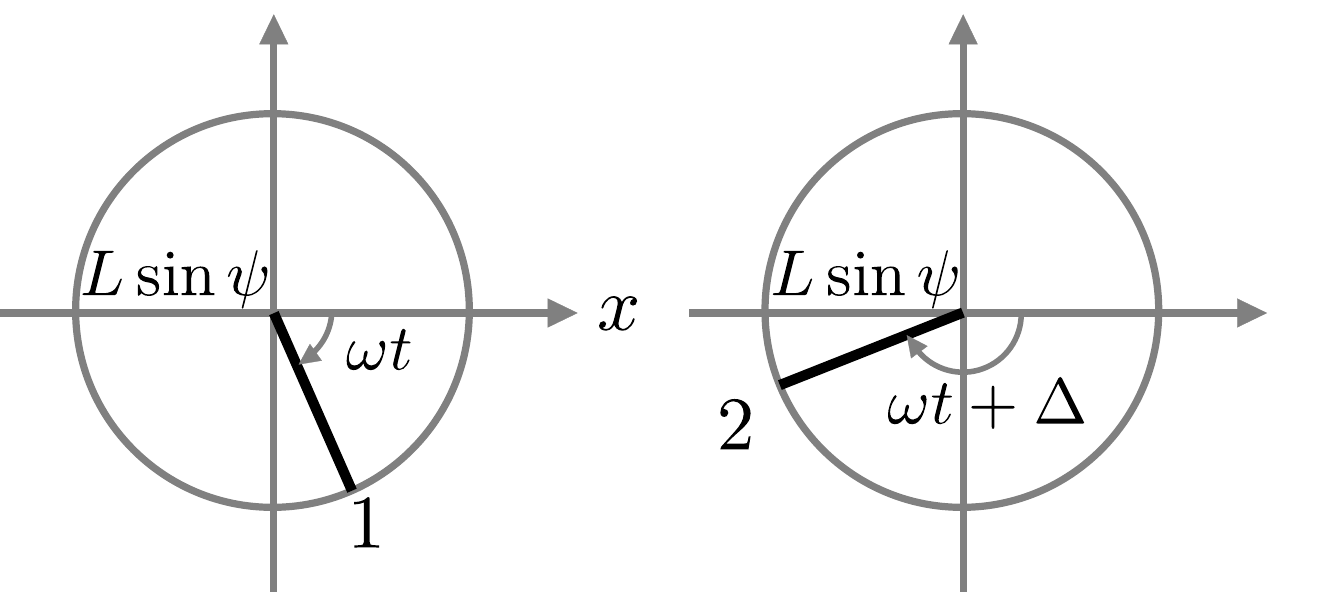}
	\caption{View of each of the orbits traced by the whirling rods in the setup of Fig.~\ref{fig:diagram_nodal}, looking down each cone axis towards the no-slip surface.}
	\label{fig:diagram_nodal_base}
\end{figure}

The setup is illustrated in Figs.~\ref{fig:diagram_nodal} and~\ref{fig:diagram_nodal_base}. 
We model each nodal cilium as a rigid rod of length $L$, a simplification that neglects slight bending of the cilia due to viscous resistance~\cite{okada2005mechanism}.
The flow outside the rods is governed by the incompressible Stokes equations for a fluid of viscosity $\mu$, Eqs.~(\ref{eq:Stokes})--(\ref{eq:incompressible}).
We consider two such rods interacting hydrodynamically in the far field. We impose their  kinematics so that each rod rotates with constant angular frequency $\omega$, sweeping out a cone of constant semi-angle $\psi$ above a no-slip surface.
The rotation is clockwise when viewed from above and the tip of each cone has a fixed position on the no-slip plane. 
Each cone axis is tilted by a constant angle $\theta$, called the ``posterior tilt", away from the normal to the no-slip surface; this tilt of nodal cilia, which is a mechanism for producing unidirectional flow, was confirmed    in experimental studies of mouse embryos~\cite{nonaka2005novo}. 
So that the rods do not make contact with the no-slip surface, we necessarily have the geometrical constraint $\theta+\psi<\pi/2$.
The two cones are separated by a distance $\ell$, assumed to be  much larger than the rod length $L$.
We impose a constant phase difference $\Delta$ between the otherwise-identical, synchronised, periodic motion of the two rods, and we  calculate below the rate of working averaged over one period as a function of the phase difference $\Delta$.

We use Cartesian coordinates $(x,y,z)$ with the no-slip surface given by $z=0$.
We choose the first cone to have its tip at the origin and the second cone to have its tip at $(\ell\cos\beta,\ell\sin\beta,0)$.
The position of a material point on the first rod is denoted by $\mathbf{r}_1(s,t)$, where $s$ is the arclength with $0\leq s\leq L$ and $t$ is time.
If the cone had no posterior tilt ($\theta=0$), then the cone axis would be aligned with the $z$ axis.
To find the cone with posterior tilt $\theta$, we rotate  the cone with zero posterior tilt, about the $x$ axis and away from the positive $y$ direction.
Thus, the position $\mathbf{r}_1(s,t)$ of the material point at arclength $s$ along the first rod at time $t$ is given by 
\begin{align}
\mathbf{r}_1(s,t)&\equiv 
\begin{pmatrix}
x_1(s,t)\\
y_1(s,t)\\
z_1(s,t)\\
\end{pmatrix}\nonumber\\
&=\begin{pmatrix}
1&0&0\\
0&\cos\theta&-\sin\theta\\
0&\sin\theta&\cos\theta\\
\end{pmatrix}
\begin{pmatrix}
s \sin\psi \cos\omega t\\
-s \sin\psi \sin\omega t\\
s \cos\psi\\
\end{pmatrix}
\nonumber \\ &= 
\begin{pmatrix}
s \sin\psi \cos\omega t\\
-s \sin\psi \sin\omega t \cos\theta - s\cos\psi\sin\theta\\
-s\sin\psi\sin\omega t\sin\theta + s \cos\psi\cos\theta\\
\end{pmatrix}.\label{eq:r1st}
\end{align}
The corresponding velocity of this material point  $\mathbf{U}_1(s,t)$ is then given by a time derivative
\begin{align}
\mathbf{U}_1(s,t)&=\frac{\partial \mathbf{r}_1(s,t)}{\partial t}
\nonumber \\ &= 
\begin{pmatrix}
-\omega s \sin\psi \sin\omega t\\
-\omega s \sin\psi \cos\omega t \cos\theta \\
-\omega s\sin\psi\cos\omega t\sin\theta\\
\end{pmatrix}.
\end{align}
Similarly, the position $\mathbf{r}_2(s,t)$ of the material point on the second rod is
\begin{align}
\mathbf{r}_2(s,t)&\equiv 
\begin{pmatrix}
x_2(s,t)\\
y_2(s,t)\\
z_2(s,t)\\
\end{pmatrix}\nonumber\\
&=
\begin{pmatrix}
s \sin\psi \cos(\omega t + \Delta)\\
-s \sin\psi \sin(\omega t + \Delta) \cos\theta - s\cos\psi\sin\theta\\
-s\sin\psi\sin(\omega t + \Delta)\sin\theta + s \cos\psi\cos\theta\\
\end{pmatrix}
+
\begin{pmatrix}
\ell\cos\beta\\
\ell\sin\beta\\
0\\
\end{pmatrix},
\end{align}
with velocity $\mathbf{U}_2(s,t)$   found as 
\begin{align}
\mathbf{U}_2(s,t)&=\frac{\partial \mathbf{r}_2(s,t)}{\partial t}
\nonumber \\ &= 
\begin{pmatrix}
-\omega s \sin\psi \sin(\omega t+\Delta)\\
-\omega s \sin\psi \cos(\omega t+\Delta) \cos\theta \\
-\omega s\sin\psi\cos(\omega t+\Delta)\sin\theta\\
\end{pmatrix}.
\end{align}

\subsection{Time-averaged rate of working}

We next use resistive-force theory~\cite{batchelor1970slender,cox70} to 
 calculate the rate at which work is done by the first rod on the fluid,  averaged over one period.  
We denote by $c_\parallel$ and $c_\perp$ the resistance coefficients of the rods in the tangential and normal directions respectively.
We approximate these coefficients as constant, even though the presence of the no-slip wall means that the coefficients may vary as the orientations of the rods change during their periodic motion~\cite{brennen77}.
Neglecting the wall effect, as in Refs.~\cite{smith2008fluid,smith2011mathematical}, will allow us to make progress analytically.

We denote by $\mathbf{F}_2(t)$ the total force exerted by the second rod on the fluid at time $t$.
To leading order in their separation, the second rod is unaffected by the first, so that using resistive-force theory,  $\mathbf{F}_2$ is given by 
\begin{align}
\mathbf{F}_2 &= \int_0^L c_\perp \mathbf{U}_2 \mathrm{\, d} s\nonumber\\
&= -\frac{1}{2} c_\perp L^2 \omega\sin\psi 
\begin{pmatrix}
\sin(\omega t + \Delta)\\
\cos(\omega t + \Delta)\cos\theta\\
\cos(\omega t + \Delta)\sin\theta\\
\end{pmatrix}.
\end{align}
Similarly to what we did in Sec.~\ref{sec:spheres}, we denote by $\mathbf{u}_{2\to 1}(s,t)$ the flow induced by the motion of the second rod, evaluated at the material point $\mathbf{r}_1(s,t)$ at arclength $s$ along the first rod, at time $t$. 
We introduce the centre of mass of the second rod $\mathbf{R}_2(t)$ with components $(X_2(t), Y_2(t), Z_2(t))$ given by 
\begin{align}
\mathbf{R}_2(t)&\equiv 
\begin{pmatrix}
X_2(t)\\
Y_2(t)\\
Z_2(t)\\
\end{pmatrix}\nonumber\\
&\equiv\mathbf{r}_2\left (\frac{L}{2},t\right )\equiv 
\begin{pmatrix}
x_2\left (\frac{L}{2},t\right )\\
y_2\left (\frac{L}{2},t\right )\\
z_2\left (\frac{L}{2},t\right )\\
\end{pmatrix}\nonumber\\
&=
\begin{pmatrix}
\frac{L}{2} \sin\psi \cos(\omega t + \Delta)\\
-\frac{L}{2} \sin\psi \sin(\omega t + \Delta) \cos\theta - \frac{L}{2}\cos\psi\sin\theta\\
-\frac{L}{2}\sin\psi\sin(\omega t + \Delta)\sin\theta + \frac{L}{2} \cos\psi\cos\theta\\
\end{pmatrix}
+
\begin{pmatrix}
\ell\cos\beta\\
\ell\sin\beta\\
0\\
\end{pmatrix}.
\end{align}
Since the rods are widely separated, we may approximate the flow $\mathbf{u}_{2\to1}$ as that due to a point force $\mathbf{F}_2$ exerted on the fluid at position $\mathbf{R}_2$. 
Using the  image system  near no-slip surfaces~\cite{blake1974fundamental},  the flow $\mathbf{u}_{2\to1}$ is given to leading order by 
\begin{align}
\mathbf{u}_{2\to1} &\simeq\frac{3 z_1Z_2 }{2\pi \mu \left \lvert \mathbf{r}_1 - \mathbf{R}_2 \right \rvert ^5 }
\begin{pmatrix}
F_{2,x}(x_1-X_2)^2+ F_{2,y}(x_1-X_2)(y_1-Y_2)\\
F_{2,x}(x_1-X_2)(y_1-Y_2)+ F_{2,y}(y_1-Y_2)^2\\
0\\
\end{pmatrix},
\end{align}
where we recall that $z_1$ is given by Eq.~(\ref{eq:r1st}).
Using this result and the approximations $x_1-X_2\simeq -\ell\cos\beta$ and $y_1-Y_2\simeq -\ell\sin\beta$ (justified since $L\ll\ell$), the flow $\mathbf{u}_{2\to1}$ is given to leading order by 
\begin{align}
\mathbf{u}_{2\to1} &\simeq -\frac{3 c_\perp L^2\omega\sin\psi z_1Z_2 [\cos\beta\sin(\omega t + \Delta)+ \sin\beta\cos(\omega t + \Delta)\cos\theta]}{4\pi \mu \ell^3 }
\begin{pmatrix}
\cos\beta\\
\sin\beta\\
0\\
\end{pmatrix}.
\end{align}

We next denote by $\mathbf{f}_1(s,t)$ the force per unit length exerted on the fluid by the material point on the first rod at arclength $s$ and time $t$. 
Then, using resistive-force theory, we have 
\begin{align}\label{eq117}
    \mathbf{f}_1(s,t) \simeq \{c_\parallel \mathbf{t}_1(t)\mathbf{t}_1(t) + c_\perp [\mathbf{1}-\mathbf{t}_1(t)\mathbf{t}_1(t)]\}\cdot[\mathbf{U}_1(s,t)-\mathbf{u}_{2\to1}(s,t)],
\end{align}
where we recall that $c_\parallel$ and $c_\perp$ are the resistance coefficients and $\mathbf{1}$ is the identity tensor. Here $\mathbf{t}_1(t)$ is  the  unit tangent vector to the first rod, given  explicitly  by
\begin{align}
    \mathbf{t}_1(t) = 
    \begin{pmatrix}
        \sin\psi\cos\omega t \\
        -\sin\psi\sin\omega t\cos\theta - \cos\psi\sin\theta\\
        -\sin\psi\sin\omega t \sin\theta + \cos\psi\cos\theta\\
    \end{pmatrix}.
\end{align}

The final term in Eq.~\eqref{eq117} captures the interaction between the two model cilia. 
Since the velocity of each rod is perpendicular to the rod's tangent, we have (omitting the $s$ and $t$ dependence for brevity)
\begin{align}
    \mathbf{f}_1 \simeq c_\perp (\mathbf{U}_1 - \mathbf{u}_{2\to1})  - (c_\parallel - c_\perp) (\mathbf{t}_1\cdot\mathbf{u}_{2\to1}) \mathbf{t}_1,
\end{align}
so that the instantaneous rate of working per unit length at each material point is 
\begin{align}
    \mathbf{f}_1 \cdot \mathbf{U}_1  \simeq c_\perp (\mathbf{U}_1 - \mathbf{u}_{2\to1})\cdot\mathbf{U}_1.
\end{align}
We can then calculate  explicitly 
\begin{align}
\mathbf{U}_1\cdot \mathbf{U}_1 = \omega^2 s^2 \sin^2\psi
\end{align}
and
\begin{align}
\mathbf{u}_{2\to1}\cdot\mathbf{U}_1\simeq&\frac{3s^2 c_\perp L^3\omega^2 \sin^2\psi  }{8\pi \mu \ell^3 }\nonumber\\
&\times(-\sin\psi\sin\omega t\sin\theta +  \cos\psi\cos\theta)[-\sin\psi\sin(\omega t + \Delta)\sin\theta +  \cos\psi\cos\theta]\nonumber\\
&\times\left [\cos\beta\sin(\omega t + \Delta)+ \sin\beta\cos(\omega t + \Delta)\cos\theta\right ]
\left (\cos\beta\sin\omega t +\sin\beta\cos\omega t\cos\theta\right).
\end{align}

From this, we find the average rate at which work is done by the first rod in one period as
\begin{align}\label{eq:123}
\frac{\omega}{2\pi}& \int_0^{{2\pi}/{\omega}} \int_0^L \mathbf{f}_1 \cdot \mathbf{U}_1 \mathrm{\, d} s \mathrm{\, d} t \nonumber\\
\simeq \frac{1}{3} c_\perp& L^3 \omega^2\sin^2\psi  \nonumber\\
\times\bigg(&1+\frac{3c_\perp L^3\sin^2\theta\sin^2\psi\left (\sin^2\beta\cos^2\theta-\cos^2\beta\right )}{64\pi\mu\ell^3} \nonumber\\
&- \frac{3 c_\perp L^3 \cos\Delta \left (2\cos^2\psi\cos^2\theta+\sin^2\psi\sin^2\theta\cos\Delta\right )\left(\sin^2\beta\cos^2\theta+\cos^2\beta\right)}{32\pi\mu \ell^3} \bigg).
\end{align}
Note that for physically relevant setups, the semi-angle of the cone $\psi$ is nonzero, and since the posterior tilt satisfies $\theta<\pi/2$, the quantity $\sin^2\beta\cos^2\theta+\cos^2\beta$ is nonzero. 
To find the corresponding result for the second rod, we simply replace the angle $\beta$ with $\pi+\beta$ and the phase difference $\Delta$ with $-\Delta$, and see (again) that the rate of working is unchanged. Thus, the total rate of energy dissipation in the fluid is  given by twice the  final result in Eq.~\eqref{eq:123}.

With this, we can now  consider the optimisation of the rate of working. 
By inspection, it is straightforward to see that in-phase motion of the rods (i.e.,~the case where $\Delta=0$) always minimises the rate of working, since the coefficients of both $\cos\Delta$ and $\cos^2\Delta$ are negative.
We may additionally compute the phase difference $\Delta$ that maximises the rate of working. 
In the case $\sin\theta=0$, it is clear that opposite-phase motion maximises the rate of working.
Otherwise, the rate of working may be written as $A - B \cos\Delta(2\cot^2\psi\cot^2\theta+\cos\Delta)$, where both $A$ and $B$ are positive  constants.
If there are no solutions to the equation 
\begin{equation}\label{eq:tosolve}
\cos\Delta=-\cot^2\psi\cot^2\theta,
\end{equation}
then opposite-phase motion of the rods ($\cos\Delta=-1$) maximises the rate of working. Luckily, for  all  physically relevant values of the parameters $\psi$ and $\theta$, there are no solutions to Eq.~\eqref{eq:tosolve}. This is due to the fact that, for a fixed value of $\psi$,  the quantity $\cot^2\psi\cot^2\theta$ decreases from $+\infty$ to $1$ as $\theta$ increases from $0$ to $\pi/2-\psi$ (which we recall is the value of $\theta$ corresponding to a cone that is tangent to the no-slip surface).
Hence, in all cases, the opposite-phase synchronised  motion of the rods maximises the rate of working, and in-phase motion minimises it. As  discussed in the next section, this compares favourably  to past theoretical predictions for  the dynamic synchronisation of   cilia~\cite{niedermayer08}.

\section{Discussion}\label{sec:discussion}

In this paper, we  investigated the energetics of synchronised motion for three different models for interacting flagella and cilia. 
In each case, we   imposed  periodic motion for the  two appendages   with a constant phase difference $\Delta$.
For two sheets waving with small amplitude and with both longitudinal and transverse modes, we found that the optimal phase difference $\Delta^*$, which minimises the rate of working of the sheets, can take all values between $-\pi$ and $\pi$ in a manner that depends on the amplitudes of and the phase difference between the waving modes, and the mean sheet separation. 
Our calculations reproduced Taylor's result that in the case of purely transverse deformation, the rate of working is minimised for in-phase waving.
We saw also that if the longitudinal mode has nonzero but small amplitude compared to the transverse mode, then the energetic minimum occurs for small, but nonzero, phase differences, reminiscent of  what is seen in the  metachronal waves of cilia and in continuum models of cilia arrays~\cite{michelin10}. Furthermore, if the sheets deform only longitudinally, then in-phase waving results in maximum dissipation, a result that we   rationalised physically.

In contrast to this, in the cases of spheres moving  along  circular orbits parallel to a no-slip plane (modelling two-stroke cilia) and whirling rods with a posterior tilt also above a no-slip surface (modelling nodal cilia), the energetic minimum always  occurs at the in-phase configuration, $\Delta=0$. 
For spheres in elliptical orbits perpendicular to a no-slip plane, the minimum rate of dissipation of energy occurs for either in-phase or opposite-phase motion, depending on the relative position and orientation of the orbits.
For all these models considered, the dissipation is maximised when the phase difference $\Delta$ differs from that giving minimum dissipation by exactly $\pi$.

The original motivation of our work was  Taylor's energetic argument as explaining the   synchronisation of spermatozoa observed in nature~\cite{taylor51}. In his paper, Taylor also remarked that the dynamics of interacting swimmers  might not necessarily result in the minimum-dissipation configuration,  depending on how the flagella are actuated. Equipped with our energetics results in the case where we imposed the kinematics, we can now examine  previous studies on the mechanisms that result in the   synchronisation of model flagella or cilia.

In an important paper in the field,  elastic compliance was proposed  as a generic mechanism allowing hydrodynamic synchronisation~\cite{niedermayer08}. 
In the model from that work, cilia are replaced by spheres, which orbit in a plane parallel to a no-slip surface and interact  hydrodynamically in the far field.
However, unlike our setup in Sec.~\ref{sec:parallelcircular}, the orbits  of the spheres are only approximately circular.
The intrinsic preferred waving motion of each cilium is modelled as a preferred radius of circular orbit and a preferred angular frequency.
An elastic restoring force then acts radially to bring the radius of the orbit back to this preferred value, and the angular frequencies of the spheres remain close to their preferred values since the spheres are widely separated. With this setup, in the case of equal intrinsic frequencies, the phase difference between the two spheres is governed by the Adler equation.  This showed remarkable agreement with experimental observation  of synchronisation of the two flagella on the unicellular biflagellate alga {\it Chlamydomonas}~\cite{goldstein2009noise}. The theoretical prediction in that case was therefore that the model  cilia synchronise to an in-phase configuration regardless of initial phase difference. In Sec.~\ref{sec:parallelcircular}, we found that for imposed constant angular velocity on precisely circular orbits,  dissipation of energy is also minimised by in-phase motion. This is also the case for the  whirling rods modelling nodal cilia found in Sec.~\ref{sec:rods}. There is  therefore full agreement in these cases between the dynamics and the energetics points of view. 

Another study also modelled cilia as spheres above a no-slip surface, but constrained them to move on fixed, tilted, precisely elliptical trajectories~\cite{vilfan06}. This led to  coupled differential equations for the evolution of the phases of the two spheres.
We recall that,  in our paper,  the angle $\beta$ describes the relative position of the two orbits   in Sec.~\ref{sec:perpendicularelliptical} for spheres in elliptical orbits and in Sec.~\ref{sec:rods} for whirling rods. In Ref.~\cite{vilfan06},    far-field calculation showed that for generic tilted orbits, the stable synchronised states are in-phase motion for  $0<\beta<\pi/2$ and $\pi<\beta<3\pi/2$, and opposite-phase motion for  $\pi/2<\beta<\pi$ and $3\pi/2<\beta<2\pi$.  These  stability results were found to also be  consistent with full numerical solutions to the dynamic equations.

We may relate the above far-field result to our result for energetics in Sec.~\ref{sec:rods} on whirling rods.
We found that in the far field, for rotation at constant angular frequency, dissipation is minimised by in-phase motion, while in Ref.~\cite{vilfan06}  the stable synchronised states are opposite-phase or in-phase motion, depending on the relative position of the two nodal cilia.
Thus, it is possible for two cilia to synchronise to a state where dissipation is maximum. 
 
We also recall our far-field result in Sec.~\ref{sec:perpendicularelliptical}, for elliptical orbits perpendicular to the no-slip surface, allowed to have different orientations.
We found that the rate of dissipation of energy is minimised either by in-phase or opposite-phase motion of the spheres, depending on the relative position and orientation of orbits. Although the setup is not quite identical to that in  Ref.~\cite{vilfan06}, we note the appearance of in-phase and opposite-phase motion in both cases. The same also occurs for interacting three-dimensional flagella, with in-phase or opposite-phase motion minimising  viscous dissipation depending on the relative position and orientation of the flagella~\cite{mettot2011energetics}.

The swimming sheet model in Sec.~\ref{sec:sheets} is the only one that we considered where the configuration with minimum energy dissipation can occur for synchronised motion that is neither in phase nor in opposite phase. Past research on dynamic models of synchronisation for swimming sheets only considered the case of transverse waving~\cite{elfring09,elfring2011passive,elfring11}. Further work will therefore be needed to probe the relationship between energetics and dynamics in the case of flexible waving swimmers with the most general swimming kinematics.

\section*{Acknowledgments}

This project has received funding   from the Faculty of Mathematics, University of Cambridge (Cambridge Summer Research in Mathematics Programme studentship to W.L.) and from the European Research Council (ERC) under the European Union's Horizon 2020 research and innovation programme (grant agreement 682754 to E.L.).

\bibliography{synch}

\end{document}